\documentclass[aps,preprint]{revtex4}%
\usepackage{amsfonts}
\usepackage{amsmath}
\usepackage{amssymb}
\usepackage{graphicx}
\usepackage{epstopdf}%
\providecommand{\U}[1]{\protect\rule{.1in}{.1in}}
\begin{document}
\preprint{ }
\title{Optical Conductivity of Topological Insulator Thin Films in a Quantizing
Magnetic Field}
\author{A. Ullah and K. Sabeeh}
\email{ksabeeh@qau.edu.pk}
\affiliation{Department of Physics, Quaid-i-Azam University, Islamabad 45320, Pakistan.}

\begin{abstract}
We determine the optical response of topological insulator thin films in the
presence of a quantizing, external magnetic field. We explicitly take into
account hybridization between the states of top and bottom surface. The
interplay between hybridization and Zeeman energies gives rise to topological
and normal insulator phases and phase transitions between them. The optical
response in the two phases and at the phase transition point is investigated.
We show that the difference in magneto-optical response can be used to
distinguish the topological phase from the normal phase of the system.
Further, the optical response also allows us to determine the gap generated by
hybridization between top and bottom surface states of topological insulator
thin films.

Pacs:

\end{abstract}
\startpage{01}
\endpage{02}
\maketitle

\section{Introduction}

Three-dimensional (3D) Topological insulators (TIs) are materials that have a
bulk bandgap but conducting surface states \cite{1,2,3,4}. These materials
usually have strong spin-orbit interaction. The conducting surface states are
protected by time-reversal symmetry. The surface states have a linear
dispersion relation and the quasiparticles (Dirac fermions) at the surface
obey the massless Dirac equation. Further, the surface states are helical
where intrinsic angular momentum (spin) and translational momentum are locked
to each other with the Dirac cone centered at the time-reversal invariant
momentum point in the Brillouin zone with spin polarized Berry
phase\cite{5,6,7}; which was confirmed by spin polarized Angle Resolved Photo
Emission Spectroscopy (ARPES). These helical Dirac fermions exist on the edge
of three-dimensional topological insulators such as $Bi_{2}Se_{3},$
$Bi_{2}Te_{3}.$

The gapless surface states are primarily responsible for transport in
topological insulators. In transport studies of TIs, a major challenge is to
separate the bulk contribution from the surface contribution. Since several
TIs are layered materials, thin films can be synthesized with the advantage
that thin films of topological insulators have minimum bulk contribution.
Experiments on thin films are being actively pursued and it has been
demonstrated that they exhibit thermoelectric effect\cite{8}, quantum spin
Hall\cite{9,10}, quantum anomalous Hall effect\cite{11} and excitonic
superfluidity\cite{12}. Additionally, thin films provide an extra tunable
degree of freedom which is their thickness. Thin films where states of top and
bottom surfaces hybridize exhibit even richer physics. This hybridization can
happen for 1-5 quintuple layers with a thickness of the order of
$5nm$\cite{11,13,14,15,16}. Fabrication of $Bi_{2}Se_{3}$ thin film by
molecular beam epitaxy\cite{13,17} as well as its low temperature transport
studies have been reported in\cite{18}. Hybridization leads to opening the gap
in the surface state dispersion\cite{19}; in other words, it provides mass to
Dirac fermions on the top and bottom surfaces. Further, this gap can be tuned
by the application of an external magnetic field. It has been shown that the
response of TI thin films in an external magnetic field is highly
nontrivial\cite{20,21,22}. TI thin films exhibit topological phases with phase
transition that can be tuned by an external magnetic field. For surface state
effects, the Fermi level has to be in the bulk band gap which can be
controlled by doping a TI\cite{5,23} or by a gate potential. Another system
that shares similarities with a TI thin film is bilayer graphene. At each
valley, there are four parabolic bands in bilayer graphene, two valence and
two conduction bands. Two of these meet at the Charge Neutrality Point (CNP)
exhibiting no energy gap in pristine bilayer; the remaining two are split from
these. Gapless bilayer is a semi-metal but that can be changed. The gap can be
opened and tuned by an electric field (gate potential) applied perpendicular
to bilayer graphene sheets. This is in contrast to a TI thin film where there
is a gap when the top and bottom surfaces are coupled and can be tuned by an
applied magnetic field.

The main question that we address in this work is the effect of hybridization
between top and bottom surface states on the magneto-optical response in thin
film TIs. For this, we determine the complex frequency dependent longitudinal
optical conductivity $\sigma(\omega);$ its real part gives the absorption as a
function of photon energy. This has been carried out for graphene which shows
good agreement between theory and experiment\cite{24,25,26,27,28}. Recently in
\cite{29} optical properties of topological insulator thin films doped with
magnetic impurities has been investigated. The authors include hybridization
effects and for exchange field that breaks time reversal symmetry show that
the value of Kerr and Faraday rotation angles is large for a wide range of
frequencies. Magneto-optical properties of TIs\cite{30} and other single layer
material such as $MoS_{2}$\cite{31,32} and silicene\cite{33} have also been
investigated. Landau levels are formed in the presence of an external magnetic
field. Transitions between the Landau levels generate absorption lines in the
magneto-optical conductivity\cite{34,35}. In \cite{36,37}, these absorption
lines were used to distinguish topological insulator phase and normal (band)
insulator phase in silicene in the presence of spin orbit interaction and
staggered potential. In this paper, we investigate topological phase
transition in thin film of a topological insulator where hybridization between
top and bottom surfaces is important. This will be done on the basis of
information obtained from magneto-optical absorption spectra. We obtain the
absorption spectra in both topological insulator phase and normal insulator
phase as well as at the Charge Neutrality Point (CNP).

This paper is organized as follows: In section 2 and 3 we develop the
theoretical model of a thin film topological insulator in a uniform external
magnetic field. In section 4, we determine the longitudinal conductivity and
transverse hall conductivity. In section 5, response to circularly polarized
light is considered. In section 6, the topological phase transition in the
semiclassical limit is investigated. In section 7 and 8, effect of broken
inversion symmetry and effect of warping in thin film topological insulator on
Landau levels has been discussed.

\section{Theory of Topological Insulator Thin Film in an External Magnetic
Field}

We consider the Hamiltonian for the surface states in a topological insulator
thin film aligned in the $xy-$plane with hybridization between the surface
states. When thin film is subjected to transverse magnetic field
$\mathbf{B=\nabla\times A}$, Landau levels with quantized energies develop in
the valence and conduction bands.\ We employ the minimal substitution
$\mathbf{p}\rightarrow$ $\mathbf{p}+\frac{e}{c}\mathbf{A}$ in the Landau gauge
for vector potential $\mathbf{A=(}0\mathbf{,}$ $xB,0)$ and $c$ is the speed of
light. The Hamiltonian of our system takes the form\cite{11}:%
\begin{equation}
\hat{H}_{\sigma\tau}=v_{f}\left[  \sigma_{x}\left(  p_{y}+\frac{eB}%
{c}x\right)  -\tau_{z}\sigma_{y}p_{x}\right]  +(\Delta_{Z}\tau_{z}+\Delta
_{H})\sigma_{z}. \label{1}%
\end{equation}
Here $(\sigma_{x},\sigma_{y},\sigma_{z})$ define Pauli matrices acting on real
spin space. $\tau_{z}=+/-$ represent the symmetric/antisymmetric linear
combination of surface states represented by $|\tau_{z}\uparrow(\downarrow
)\rangle=1/\sqrt{2}(|t\uparrow(\downarrow)\rangle+\tau_{z}|b\uparrow
(\downarrow)\rangle)$\cite{11}. Here $t$ represents the top surface and $b$
the bottom surface of the thin film. $v_{f}$ is the Fermi velocity of Dirac
fermions on the surface. Moreover, we have Zeeman energy $\Delta_{Z}=g\mu
_{B}B/2,$ the effective Lande factor $g,$ the Bohr magneton $\mu_{B}$, and
$\Delta_{H}$ represents the hybridization contribution which is due to the
hybridization between upper and lower surfaces of the TI. As $p_{x}$ and $x$
do not commute, we can write the Hamiltonian in terms of dimensionless
operators%
\begin{equation}
\hat{H}_{\sigma\tau}=\frac{v_{f}}{l_{B}}\left[  \sigma_{x}l_{B}\hat{P}%
+\tau_{z}\sigma_{y}\frac{\hat{Q}}{l_{B}}\right]  +(\Delta_{Z}\tau_{z}%
+\Delta_{H})\sigma_{z},
\end{equation}
where $l_{B}=\sqrt{c/eB}$ is the magnetic length. $\hat{Q}=-l_{B}^{2}p_{x}$
and $\hat{P}=p_{y}+\frac{eB}{c}x$ such that $[\hat{Q},\hat{P}]=i\hbar$.
Employing the ladder operators $a=1/\sqrt{2\hbar}l_{B}(\hat{Q}+il_{B}^{2}%
\hat{P})$ and $a^{\dagger}=1/\sqrt{2\hbar}l_{B}(\hat{Q}-il_{B}^{2}\hat{P}),$
we may express the Hamiltonian as
\begin{equation}
\hat{H}_{\sigma\tau}=\sqrt{\frac{\hbar}{2}}\frac{v_{f}}{l_{B}}(i\sigma
_{x}(a^{\dagger}-a)+\tau_{z}\sigma_{y}(a+a^{\dagger}))+(\Delta_{Z}\tau
_{z}+\Delta_{H})\sigma_{z}. \label{x}%
\end{equation}
which can also be written as
\begin{equation}
\hat{H}_{\tau_{z}=+1}=\left(
\begin{array}
[c]{cc}%
(\Delta_{Z}+\Delta_{H}) & -i\sqrt{2\hbar}\frac{v_{f}}{l_{B}}a\\
i\sqrt{2\hbar}\frac{v_{f}}{l_{B}}a^{\dagger} & -(\Delta_{Z}+\Delta_{H})
\end{array}
\right)  \label{b}%
\end{equation}%
\begin{equation}
\hat{H}_{\tau_{z}=-1}=\left(
\begin{array}
[c]{cc}%
-(\Delta_{Z}-\Delta_{H}) & i\sqrt{2\hbar}\frac{v_{f}}{l_{B}}a^{\dagger}\\
-i\sqrt{2\hbar}\frac{v_{f}}{l_{B}}a & (\Delta_{Z}-\Delta_{H})
\end{array}
\right)  \label{c}%
\end{equation}

The energy of the Landau levels (LLs) is given by
\begin{equation}
E_{n}^{\tau_{z}}=sgn(n)\sqrt{2\hbar v_{f}^{2}eB\left\vert n\right\vert
+(\Delta_{Z}+\tau_{z}\Delta_{H})^{2}}, \label{e}%
\end{equation}%
\begin{equation}
E_{0}^{\tau_{z}}=-(\Delta_{Z}+\tau_{z}\Delta_{H}).
\end{equation}
$\omega_{B}=v_{f}/l_{B}$ is the cyclotron frequency of Dirac fermions.
$n=0,\pm1,\pm2,....$ is the Landau level index. An important feature of the
energy spectrum is the splitting of $n\neq0$ Landau levels for non-zero value
of Zeeman energy and hybridization between top and bottom surface states. This
splitting of $n\neq0$ requires both Zeeman energy and hybridization to be
nonzero. Further, the energy spectrum is electron-hole symmetric in the
absence of Zeeman energy $(\Delta_{Z}=0).$ For $\Delta_{z}<\Delta_{H}$ with a
finite hybridization gap, it is not strictly electron-hole symmetric; the
$n\neq0$ spectrum maintains this symmetry where as $n=0$ spectrum does not.
Note that a quadratic term can appear in the Hamiltonian even in the absence
of both Zeeman energy and hybridization\cite{38}, if there is no electron-hole
symmetry, as shown by angle resolved photoemission spectroscopy. In our case
we have not considered a quadratic term in Eq. (\ref{1}) as it can be
neglected when the system is doped such that the Dirac point is close to the
charge neutrality point (CNP), which is the focus of our work. The $n=0$
Landau level splits only when $\Delta_{H}$ is nonzero. The LL energy spectrum
carries important information regarding topological phase transition in the
system. The $n=0$ Landau level $E_{0}^{-}$ changes sign during the phase
transition from normal insulator $(\Delta_{z}<\Delta_{h})$ to topological
insulator $(\Delta_{z}>\Delta_{h})$. For normal insulator phase $E_{0}^{-}$ is
hole like and for topological insulator phase it is electron like. This
represents an extra filled Landau level which gives rise to Hall conductivity
$e^{2}/h$\cite{20}; hence we can write%
\begin{equation}
\sigma_{xy}=\frac{e^{2}}{2h}(sgn(\Delta_{Z}-\Delta_{H})+1).
\end{equation}
At $\Delta_{Z}=\Delta_{H}$, $E_{0}^{-}$ has exactly zero energy and is at the
charge neutrality point (CNP). If the chemical potential is tuned to CNP, this
zeroth Landau level will be partially filled. The plot of Landau levels with
respect to magnetic field is shown in Fig. 1 with $g=60$ and $\mu
_{B}=5.788\times10^{-6}\frac{eV}{T}.$ Similar to graphene all $n\neq0$ Landau
levels scale as $\sqrt{B}$. But unlike graphene $n=0$ Landau levels do not sit
at zero energy when $\Delta_{Z}\neq\Delta_{H}$. The energy of one of the $n=0$
Landau levels becomes zero for $\Delta_{Z}=\Delta_{H}$. At that point
$\Delta_{H}=4meV$ with magnetic field $B=2.3T$. This represents the
topological phase as shown in Fig. 1.

Using Eq. (\ref{b}) and Eq. (\ref{c}), the eigenvectors for symmetric surface
states are
\begin{equation}
|\tilde{n}\rangle_{\tau_{z}=+1}=%
\begin{pmatrix}
A_{n}|\left\vert n\right\vert -1\rangle\\
B_{n}|\left\vert n\right\vert \rangle
\end{pmatrix}
,
\end{equation}

and%
\[
|\tilde{n}\rangle_{\tau_{z}=-1}=%
\begin{pmatrix}
A_{n}|\left\vert n\right\vert \rangle\\
B_{n}|\left\vert n\right\vert -1\rangle
\end{pmatrix}
,
\]

where $|\left\vert n\right\vert \rangle$ is an orthonormal Fock state of the
harmonic oscillator and%
\begin{equation}
A_{n}=\{%
\begin{array}
[c]{c}%
\frac{1}{\sqrt{2}}\left(  1+sgn(n)\frac{(\Delta_{Z}+\tau_{z}\Delta_{H}%
)}{\left\vert E_{n}^{\tau_{z}}\right\vert }\right)  ^{1/2},\text{ }n\neq0,\\
0,\text{
\ \ \ \ \ \ \ \ \ \ \ \ \ \ \ \ \ \ \ \ \ \ \ \ \ \ \ \ \ \ \ \ \ \ \ \ \ }%
n=0,
\end{array}
\end{equation}

and%
\begin{equation}
B_{n}=\{%
\begin{array}
[c]{c}%
\frac{1}{\sqrt{2}}\left(  1-sgn(n)\frac{(\Delta_{Z}+\tau_{z}\Delta_{H}%
)}{\left\vert E_{n}^{\tau_{z}}\right\vert }\right)  ^{1/2},\text{ }n\neq0,\\
1,\text{
\ \ \ \ \ \ \ \ \ \ \ \ \ \ \ \ \ \ \ \ \ \ \ \ \ \ \ \ \ \ \ \ \ \ \ }n=0.
\end{array}
\end{equation}

\section{Density of States}

To shed further light on the energy spectrum of our system, we determine the
Dirac fermion density of states. The Green's function associated with our
Hamiltonian is
\begin{equation}
G(\omega,n,\tau_{z})=\underset{\tau_{z}}{\sum}\frac{1}{\omega-sgn(n)\sqrt
{2v_{f}^{2}\hbar eB\left\vert n\right\vert +(\Delta_{Z}+\tau_{z}\Delta
_{H})^{2}}+i\eta}%
\end{equation}
From which we can compute the density of states $D(\omega)$ as%
\begin{equation}
D(\omega)=-\frac{1}{2\pi^{2}l_{B}}[\underset{n=-\infty}{\overset{\infty}{\sum
}}\underset{\tau_{z}}{\sum}\operatorname{Im}G(\omega,n,\tau_{z})]
\end{equation}
which can be expressed as%
\begin{equation}
D(\omega)=\frac{-1}{\pi}\frac{1}{2\pi l_{B}^{2}}[\underset{n\neq0}%
{\underset{n=-\infty}{\overset{\infty}{\sum}}}\underset{\tau_{z}}{\sum
}\operatorname{Im}G(\omega,n,\tau_{z})+\underset{\tau_{z}}{\sum}%
\operatorname{Im}G(\omega,0,\tau_{z})].
\end{equation}
This yields
\begin{equation}
D(\omega)=\frac{1}{2\pi l_{B}^{2}}[\underset{n\neq0}{\underset{n=-\infty
}{\overset{\infty}{\sum}}}\underset{\tau_{z}}{\sum}\delta(\omega
-sgn(n)\sqrt{2v_{f}^{2}\hbar eB\left\vert n\right\vert +(\Delta_{Z}+\tau
_{z}\Delta_{H})^{2}})+\underset{\tau_{z}}{\sum}\delta(\omega+(\Delta_{Z}%
+\tau_{z}\Delta_{H}))
\end{equation}

The plot of density of states $D(\omega)$ is shown in Fig. 2 as a function of
energy$.$ We used $B=1T$ and $\eta=0.12\Delta_{H}$; $\eta$ is the scattering
rate which results in broadening of the states. The two $n=0$ Landau levels
are located at $\omega=-(\Delta_{Z}+\Delta_{H})$ and $\omega=-(\Delta
_{Z}-\Delta_{H})$. At $\Delta_{Z}=\Delta_{H}$ we have peak at CNP representing
partial filled Landau level. Peak at CNP shifts to the hole region by
increasing Zeeman energy or it shifts to the electron region by decreasing
Zeeman energy relative to the hybridization as shown in Fig. 2(a) and 2(c).

\section{Magneto-Optical Conductivity}

We determine the magneto-optical conductivity with in the linear response
regime using the Kubo formula\cite{28,39}%

\begin{equation}
\sigma_{\alpha\beta}=\frac{i}{2\pi l_{B}^{2}}\underset{\tau_{z}=\pm1}{%
%TCIMACRO{\dsum }%
%BeginExpansion
{\displaystyle\sum}
%EndExpansion
}\underset{nm}{\sum}\frac{f_{m}-f_{n}}{(\varepsilon_{n}-\varepsilon_{m})}%
\frac{\langle\tilde{m}|\hat{\jmath}_{\alpha}|\tilde{n}\rangle\langle\tilde
{n}|\hat{\jmath}_{\beta}|\tilde{m}\rangle}{\hbar\omega+\varepsilon
_{m}-\varepsilon_{n}+i\eta},
\end{equation}
where $\hat{\jmath}_{\alpha}=e\frac{\partial H}{\partial k_{\alpha}}$ and
$f_{m}=1/[1+\exp(\beta(\varepsilon_{m}-\mu))]$ is the Fermi distribution
function with $\beta=1/k_{B}T,$ $\varepsilon_{m}$ is the energy of $m$th
Landau level and $\eta$ is scattering rate taken as constant. We will take
states $m$ to be occupied and $n$ as unoccupied LLs. The selection rule for
Landau levels transition is $\left\vert n\right\vert =\left\vert m\right\vert
\pm1$ determined by the evaluation of matrix elements. At zero temperature we
can drop the absolute value of $n$ and all transition to negative Landau
levels are Pauli blocked. For longitudinal magneto-optical conductivity with
$\hat{\jmath}_{x}=ev_{f}(-\tau_{z}\sigma_{y}),$ the matrix element for
symmetric eigenstates is determined as%
\begin{equation}
\langle\left\vert \tilde{m}\right\vert |\hat{\jmath}_{x}|\tilde{n}%
\rangle\langle\tilde{n}|\hat{\jmath}_{x}|\left\vert \tilde{m}\right\vert
\rangle=v_{f}^{2}e^{2}[(A_{m}B_{n})^{2}\delta_{\left\vert m\right\vert
-1,n}+(A_{n}B_{m})^{2}\delta_{\left\vert m\right\vert +1,n}],
\end{equation}
\bigskip and for antisymmetric eigenstates, it is%
\begin{equation}
\langle\left\vert \tilde{m}\right\vert |\hat{\jmath}_{x}|\tilde{n}%
\rangle\langle\tilde{n}|\hat{\jmath}_{x}|\left\vert \tilde{m}\right\vert
\rangle=v_{f}^{2}e^{2}[(A_{m}B_{n})^{2}\delta_{\left\vert m\right\vert
-1,n}+(A_{n}B_{m})^{2}\delta_{\left\vert m\right\vert +1,n}].
\end{equation}
Therefore, we obtain%
\begin{equation}
\frac{\sigma_{xx}}{\sigma_{o}}=\frac{2iv_{f}^{2}e\hbar B}{\pi}\underset
{\tau_{z}=\pm1,mn}{\sum}\frac{[(A_{m}B_{n})^{2}\delta_{n,\left\vert
m\right\vert -1}+(A_{n}B_{m})^{2}\delta_{n,\left\vert m\right\vert +1}%
]}{(E_{n}^{\tau_{z}}-E_{m}^{\tau_{z}})(\hbar\omega+E_{m}^{\tau_{z}}%
-E_{n}^{\tau_{z}}+i\eta)}.
\end{equation}
where $\sigma_{o}=e^{2}/4\hbar.$ From the above result it is clear that for
possible transitions we must have $n=\left\vert m\right\vert \pm1$. Using the
selection rule $n=\left\vert m\right\vert \pm1$ we can write%
\begin{align}
\frac{\sigma_{xx}(\omega)}{\sigma_{o}}  &  =\frac{2iv_{f}^{2}e\hbar B}{\pi
}\underset{\tau_{z}=\pm1,m}{\sum}[\frac{(A_{\left\vert m\right\vert +1}%
B_{m})^{2}}{(E_{\left\vert m\right\vert +1}^{\tau_{z}}-E_{m}^{\tau_{z}}%
)(\hbar\omega+E_{m}^{\tau_{z}}-E_{\left\vert m\right\vert +1}^{\tau_{z}}%
+i\eta)}+\nonumber\\
&  \frac{(A_{m}B_{\left\vert m\right\vert -1})^{2}}{(E_{\left\vert
m\right\vert -1}^{\tau_{z}}-E_{m}^{\tau_{z}})(\hbar\omega+E_{m}^{\tau_{z}%
}-E_{\left\vert m\right\vert -1}^{\tau_{z}}+i\eta)}]
\end{align}%
\begin{align}
\frac{\operatorname{Re}\sigma_{xx}(\omega)}{\sigma_{o}}  &  =\frac{2v_{f}%
^{2}e\hbar B}{\pi}\underset{\tau_{z}=\pm1,m}{\sum}[\frac{[(A_{\left\vert
m\right\vert +1}B_{m})^{2}\times\eta}{(E_{\left\vert m\right\vert +1}%
^{\tau_{z}}-E_{m}^{\tau_{z}})[(\hbar\omega+E_{m}^{\tau_{z}}-E_{\left\vert
m\right\vert +1}^{\tau_{z}})^{2}+\eta^{2}]}+\nonumber\\
&  \frac{(A_{m}B_{\left\vert m\right\vert -1})^{2}\times\eta}{(E_{\left\vert
m\right\vert -1}^{\tau_{z}}-E_{m}^{\tau_{z}})[(\hbar\omega+E_{m}^{\tau_{z}%
}-E_{\left\vert m\right\vert -1}^{\tau_{z}})^{2}+\eta^{2}]}]
\end{align}

For transverse Hall conductivity $\hat{\jmath}_{x}=ev_{f}(-\tau_{z}\sigma
_{y})$ and $\hat{\jmath}_{y}=ev_{f}(\sigma_{x})$. The matrix elements are
evaluated to yield
\begin{equation}
\frac{\sigma_{xy}}{\sigma_{o}}=\frac{2v_{f}^{2}e\hbar B}{\pi}\underset
{\tau_{z}=\pm1,mn}{\sum}\frac{\tau_{z}[(A_{n}B_{m})^{2}\delta_{n,\left\vert
m\right\vert +1}-(A_{m}B_{n})^{2}\delta_{n,\left\vert m\right\vert -1}%
)]}{(E_{n}^{\tau_{z}}-E_{m}^{\tau_{z}})(\hbar\omega+E_{m}^{\tau_{z}}%
-E_{n}^{\tau_{z}}+i\eta)}%
\end{equation}

\bigskip Using selection rule $n=\left\vert m\right\vert \pm1$ we can write%
\begin{align}
\frac{\sigma_{xy}}{\sigma_{o}}  &  =\frac{2v_{f}^{2}e\hbar B}{\pi}%
\underset{\tau_{z}=\pm1,m}{\sum}[\frac{\tau_{z}[(A_{\left\vert m\right\vert
+1}B_{m})^{2}]}{(E_{\left\vert m\right\vert +1}^{\tau_{z}}-E_{m}^{\tau_{z}%
})(\hbar\omega+E_{m}^{\tau_{z}}-E_{\left\vert m\right\vert +1}^{\tau_{z}%
}+i\eta)}-\nonumber\\
&  \frac{\tau_{z}(A_{m}B_{\left\vert m\right\vert -1})^{2}}{(E_{\left\vert
m\right\vert -1}^{\tau_{z}}-E_{m}^{\tau_{z}})(\hbar\omega+E_{m}^{\tau_{z}%
}-E_{\left\vert m\right\vert -1}^{\tau_{z}}+i\eta)}]
\end{align}
This is the general expression for transverse Hall conductivity representing
transition from $m$ to $\left\vert m\right\vert \pm1$ state. We can also
determine $\operatorname{Im}\sigma_{xy}/\sigma_{o}$ as
\begin{align}
\frac{\operatorname{Im}\sigma_{xy}}{\sigma_{o}}  &  =\frac{2v_{f}^{2}e\hbar
B}{\pi}\underset{\tau_{z}=\pm1,m}{\sum}[\frac{-\tau_{z}(A_{\left\vert
m\right\vert +1}B_{m})^{2}\times\eta}{(E_{\left\vert m\right\vert +1}%
^{\tau_{z}}-E_{m}^{\tau_{z}})[(\hbar\omega+E_{m}^{\tau_{z}}-E_{\left\vert
m\right\vert +1}^{\tau_{z}})^{2}+\eta^{2}]}+\nonumber\\
&  \frac{\tau_{z}(A_{m}B_{\left\vert m\right\vert -1})^{2}\times\eta
}{(E_{\left\vert m\right\vert -1}^{\tau_{z}}-E_{m}^{\tau_{z}})[(\hbar
\omega+E_{m}^{\tau_{z}}-E_{\left\vert m\right\vert -1}^{\tau_{z}})^{2}%
+\eta^{2}]}]. \label{d}%
\end{align}
Fig. $3(c)$ shows $\operatorname{Re}\sigma_{xx}(\omega)/\sigma_{o}$ as a
function of frequency in normal insulator phase showing absorption line for
interband transitions with $v_{f}^{2}e\hbar B=1.6\times10^{-4}$ for magnetic
field of $1$ Tesla and $\mu=0$. The transition energy is determined from the
energy gap between Landau levels satisfying the selection rule for allowed
transitions. The first two absorption peaks correspond to $E_{-1}%
^{-}\rightarrow E_{0}^{-}$ and $E_{0}^{+}\rightarrow E_{1}^{+}$ transitions.
These transitions involve zeroth Landau level. The energy of first peak is
$E_{0}^{-}-E_{-1}^{-}$ and for the second peak it is $E_{1}^{+}-E_{0}^{+}$.
Each of these peaks represents single transition. The absorption peaks for
allowed transitions which involve Landau levels other then $E_{0}^{\tau_{z}}$
represent the sum of absorption peaks of two transitions in the absence of
hybridization($\Delta_{H}=0$), one transition for $E_{-n}\rightarrow E_{n+1}$
and another transition for $E_{-(n+1)}\rightarrow E_{n}$. However, for finite
hybridization $\Delta_{H},$ each peak splits into two peaks for $\tau_{z}=+1$
and $\tau_{z}=-1.$ First peak represents $E_{-n}^{-1}\rightarrow E_{n+1}^{-1}$
and $E_{-(n+1)}^{-1}\rightarrow E_{n}^{-1}$ transitions. The second peak
represents $E_{-n}^{+1}\rightarrow E_{n+1}^{+1}$ and $E_{-(n+1)}%
^{+1}\rightarrow E_{n}^{+1}$transitions. For a fixed hybridization, the
spliting between these peaks depends on the applied magnetic field; the energy
gap is $(\Delta_{Z}+\tau_{z}\Delta_{H}).$ Further, the spacing between
absorption peaks also depends on the broadening parameter $\eta;$ we have
taken its value to be $\eta=0.15\Delta_{H}$ estimated from experimental
findings\cite{40}\textbf{.} Moreover, at low magnetic fields, in the NI phase,
the splitting between $\tau_{z}=-1$ and $\tau_{z}=-1$ is very small as shown
in Fig. $3(c)$. Fig. $3(b)$ shows the real part of $\sigma_{xx}(\omega)$ at
CNP. For the 1st peak two transitions, represented by arrows, $E_{0}%
^{-}\rightarrow E_{1}^{-}$ and $E_{-1}^{-}\rightarrow E_{0}^{-}$ contribute.
While 2nd peak represents $E_{0}^{+}\rightarrow E_{1}^{+}$ transition. The
value of Zeeman interaction is large enough that it can open a gap between LLs
of different $\tau_{z}$ but same Landau index $n$ resulting in splitting of
absorption peaks; this is clearly seen in 3rd peak. Fig. 3(a) shows
$\operatorname{Re}\sigma_{xx}(\omega)/\sigma_{o}$ as a function of frequency
in the topological insulator phase with broken particle-hole symmetry for
interband transitions. The first absorption peak represents the transition
$E_{0}^{-}\rightarrow E_{1}^{-}$, while the second peak represents the
$E_{0}^{+}\rightarrow E_{1}^{+}$ transition. An important feature of the
absorption spectra is that the $E_{0}^{-}\rightarrow E_{1}^{-}$ has replaced
the $E_{-1}^{-}\rightarrow E_{0}^{-}$ transition which was allowed in normal
insulator phase but Pauli blocked in topological insulator phase. Another
difference arises in absorption peaks for topological insulator phase when
$\Delta_{Z}$ is large. The large value of $\Delta_{Z}$ induces significant gap
between $\tau_{z}=+1$ and $\tau_{z}=-1$ Landau levels of same $n$. The effect
of this gap can be seen in the absorption peaks for $\operatorname{Re}%
\sigma_{xx}(\omega)/\sigma_{o}$. The splitting in the peaks is significant as
compared to normal insulator phase of same $\tau_{z}$. Each peak in the pair
has same transition energy for the transition $m\rightarrow\left\vert
m\right\vert \pm1$ with same $\tau_{z}$. For example, the third absorption
peak is the sum of two peaks resulting from the $E_{-1}^{-}\rightarrow
E_{2}^{-}$ and $E_{-2}^{-}\rightarrow E_{1}^{-}$ transition with same energy.
Similarly the fourth peak represents sum of $E_{-1}^{+}\rightarrow E_{2}^{+}$
and $E_{-2}^{+}\rightarrow E_{1}^{+}$ transitions.

The absorption peaks obtained from $\operatorname{Im}\sigma_{xy}%
(\omega)/\sigma_{o}$ have significant differences for the two phases. Fig.
$4(c)$ represents the absorption peak for normal insulator phase. The first
two peaks represent the absorption peaks for $\operatorname{Im}\sigma
_{xy}(\omega)/\sigma_{o}$ resulting from the $E_{-1}^{-}\rightarrow E_{0}^{-}$
and $E_{0}^{+}\rightarrow E_{1}^{+}$ transitions. It shows same behavior as
for Re$\sigma_{xx}(\omega)/\sigma_{o}$ in the normal insulator phase. The
other peaks show different behavior. In Eq. (\ref{d}) for $\operatorname{Im}%
\sigma_{xy}(\omega)/\sigma_{o}$ the transition $m\rightarrow\left\vert
m\right\vert +1$ with $\tau_{z}=+1$ has positive amplitude and for
$m\rightarrow\left\vert m\right\vert -1$ with $\tau_{z}=+1$ the amplitude is
negative. Similarly for $\tau_{z}=-1$ the transition $m\rightarrow\left\vert
m\right\vert +1$ and $m\rightarrow\left\vert m\right\vert -1$ has negative and
positive amplitudes respectively. These differences not only decrease the
height of absorption peaks but they also create oscillations in absorption
peaks for $\operatorname{Im}\sigma_{xy}(\omega)/\sigma_{o}$ in the topological
insulator phase. For example the terms for the first peak in Eq. (\ref{d}) for
$\tau_{z}=-1$ have $E_{-1}^{-}\rightarrow E_{2}^{-}$ and $E_{-2}%
^{-}\rightarrow E_{1}^{-}$ transitions. Both of these transitions have
opposite signs but have same transition energy. So these terms decrease the
height of absorption peaks. Similar behavior is seen for $\tau_{z}=+1$. The
amplitude for the transition with $m\rightarrow\left\vert m\right\vert +1$
with $\tau_{z}=+1$ will always be greater than the amplitude of transition
$m\rightarrow\left\vert m\right\vert -1$ for $\tau_{z}=-1$ in both topological
insulator and normal insulator phase. At CNP, the contribution to absorption
peak in $\operatorname{Im}\sigma_{xy}(\omega)/\sigma_{o}$ resulting from
transitions involving $\tau_{z}=-1$ LLs are absent. At $\Delta_{Z}=\Delta_{H}$
all transitions resulting from $\tau_{z}=-1$ cancel out and the contribution
to absorption peaks is given by transition between Landau levels with
$\tau_{z}=+1$. The first peak in Fig. $4(b)$ represents the transition
$E_{0}^{+}\rightarrow E_{1}^{+}$ while for the 2nd case two transitions
contribute $i.e.$ $E_{-1}^{+}\rightarrow E_{2}^{+}$ and $E_{-2}^{+}\rightarrow
E_{1}^{+},$ transitions. Fig. $4(a)$ represents the absorption peaks for
$\operatorname{Im}\sigma_{xy}(\omega)/\sigma_{o}$ in TI phase. At $\Delta
_{Z}>\Delta_{H}$ the absorption peaks has negative and positive peaks. The
first negative peak results from the transition $E_{0}^{-}\rightarrow
E_{1}^{-}$, while the second positive peak represent the $E_{0}^{+}\rightarrow
E_{1}^{+}$ transition.

A schematic diagram which helps us to understand the behavior of the
absorption lines that we have described is shown in Fig. 5 for NI, in Fig. 6
at CNP and in Fig. 7 in TI phase. On the left side we shown the Landau index
$n$ with energy define by Eq. (\ref{e}). The blue lines represent the LLs with
$\tau_{z}=-1$ while red lines represent LLs for $\tau_{z}=+1$. The bold black
line gives the possible values of chemical potential $\mu=0$. The possible
optical transitions are indicated by vertical arrows and they connect the
levels $m$ to $\left\vert m\right\vert \pm1$ only. Moving from left to right
in Fig. 5 in NI phase we see first two single transition with different
transition energy from $E_{-1}^{-}$ to $E_{0}^{-}$ and $E_{0}^{+}$ to
$E_{1}^{+}$, then a pair of interband transitions from $E_{-1}^{-}$ to
$E_{2}^{-}$ and $E_{-2}^{-}$ to $E_{1}^{-}$ followed by another pair
$E_{-1}^{+}$ to $E_{2}^{+}$ and $E_{-2}^{+}$ to $E_{1}^{+}$. The difference
between the transition energy of these two pairs is very small in NI phase
showing small spliting in absorption peak of Re$\sigma_{xx}(\omega)/\sigma
_{o}$. For TI the schematic of allowed transitions is shown in Fig. 7. The
main difference in the TI and NI phase arises in the first transition. The
transition $E_{-1}^{-}$ to $E_{0}^{-}$ in NI is replaced by the transition
$E_{0}^{-}$ to $E_{1}^{-}$ in TI. The $2nd$ transition is from $E_{0}^{+}$ to
$E_{1}^{+}.$ It is followed by a pair of transitions from $E_{-1}^{-}$ to
$E_{2}^{-}$ and $E_{-2}^{-}$ to $E_{1}^{-}$ followed by another pair
$E_{-1}^{+}$ to $E_{2}^{+}$ and $E_{-2}^{+}$ to $E_{1}^{+}$. The absorption
peaks of the two pairs are well separated in the response function of
Re$\sigma_{xx}(\omega)/\sigma_{o}$ as shown in Fig. 3(a). At CNP, the allowed
transition are shown in Fig. 6. The first two transition involve partially
filled Landau level $E_{0}^{-}$. These transition are $E_{0}^{-}$ to
$E_{1}^{-}$ and $E_{-1}^{-}$ to $E_{0}^{-}$. These have same transition
energy. These transitions are followed by $E_{0}^{+}$ to $E_{1}^{+}$, then
pair of transitions same as describe previously. Fig. 8 and Fig. 9 represent
Re$\sigma_{xx}(\omega)/\sigma_{o}$ resulted from allowed transitions in NI and
TI phases respectively for $\mu=0.02eV$. The red peak represents the
absorption lines contributed by intraband transitions. While the black peaks
represent the allowed interband transition splited in TI phase at high
magnetic field. A schematic diagram for allowed transition with nonzero value
chemical potential is shown in Fig. 10 for low magnetic field (NI) and in Fig.
11 for high magnetic field (TI) showing inter and intra band transitions.

\section{Circularly-Polarized Light}

For circularly Polarized light the conductivity is written as $\sigma
_{xx}(\omega)\pm i\sigma_{xy}(\omega)$ with $(+)$ representing the right
handed polarization and $(-)$ representing left handed polarization. The
circularly polarized light shows different behavior in normal insulator and in
topological insulator phase. The absorptive part of conductivity is%
\begin{equation}
\operatorname{Re}\sigma_{\pm}=\operatorname{Re}\sigma_{xx}(\omega
)\mp\operatorname{Im}\sigma_{xy}(\omega)
\end{equation}
For normal insulator%
\begin{align}
\frac{\operatorname{Re}\sigma_{+}(\omega)}{\sigma_{o}}  &  =\frac{2v_{f}%
^{2}e\hbar B}{\pi}\underset{m=1}{\overset{\infty}{\sum}}[\frac{[(A_{\left\vert
m\right\vert +1}B_{m})^{2}\times\eta}{(E_{\left\vert m\right\vert +1}%
^{-}-E_{m}^{-})(\hbar\omega+E_{m}^{-}-E_{\left\vert m\right\vert +1}^{-}%
)^{2}+\eta^{2})}+\\
&  \frac{(A_{m}B_{\left\vert m\right\vert -1})^{2}\times\eta}{(E_{\left\vert
m\right\vert -1}^{+}-E_{m}^{+})(\hbar\omega+E_{m}^{+}-E_{\left\vert
m\right\vert -1}^{+})^{2}+\eta^{2})}]\nonumber
\end{align}

and
\begin{align}
\operatorname{Re}\sigma_{-}(\omega)  &  =\frac{2v_{f}^{2}e\hbar B}{\pi
}\underset{m=0}{\overset{\infty}{\sum}}[\frac{[(A_{\left\vert m\right\vert
+1}B_{m})^{2}\times\eta}{(E_{\left\vert m\right\vert +1}^{+}-E_{m}^{+}%
)(\hbar\omega+E_{m}^{+}-E_{\left\vert m\right\vert +1}^{+})^{2}+\eta^{2}%
)}+\nonumber\\
&  \frac{(A_{m}B_{\left\vert m\right\vert -1})^{2}\times\eta}{(E_{\left\vert
m\right\vert -1}^{-}-E_{m}^{-})(\hbar\omega+E_{m}^{-}-E_{\left\vert
m\right\vert -1}^{-})^{2}+\eta^{2})}]
\end{align}

For topological insulator%
\begin{align}
\frac{\operatorname{Re}\sigma_{+}(\omega)}{\sigma_{o}}  &  =\frac{2v_{f}%
^{2}e\hbar B}{\pi}[\underset{m=0}{\overset{\infty}{\sum}}\left(
\frac{[(A_{\left\vert m\right\vert +1}B_{m})^{2}\times\eta}{(E_{\left\vert
m\right\vert +1}^{-}-E_{m}^{-})(\hbar\omega+E_{m}^{-}-E_{\left\vert
m\right\vert +1}^{-})^{2}+\eta^{2})}\right)  +\\
&  \underset{m=1}{\overset{\infty}{\sum}}\left(  \frac{(A_{m}B_{\left\vert
m\right\vert -1})^{2}\times\eta}{(E_{\left\vert m\right\vert -1}^{+}-E_{m}%
^{+})(\hbar\omega+E_{m}^{+}-E_{\left\vert m\right\vert -1}^{+})^{2}+\eta^{2}%
)}\right)  ]\nonumber
\end{align}

\bigskip and%

\begin{align}
\frac{\operatorname{Re}\sigma_{-}(\omega)}{\sigma_{o}}  &  =\frac{2v_{f}%
^{2}e\hbar B}{\pi}[\underset{m=0}{\overset{\infty}{\sum}}\left(
\frac{[(A_{\left\vert m\right\vert +1}B_{m})^{2}\times\eta}{(E_{\left\vert
m\right\vert +1}^{+}-E_{m}^{+})(\hbar\omega+E_{m}^{+}-E_{\left\vert
m\right\vert +1}^{+})^{2}+\eta^{2})}\right)  +\\
&  \underset{m=1}{\overset{\infty}{\sum}}\left(  \frac{(A_{m}B_{\left\vert
m\right\vert -1})^{2}\times\eta}{(E_{\left\vert m\right\vert -1}^{-}-E_{m}%
^{-})(\hbar\omega+E_{m}^{-}-E_{\left\vert m\right\vert -1}^{-})^{2}+\eta^{2}%
)}\right)  ]\nonumber
\end{align}
Fig. 12 shows absorption peaks for right handed circularly polarized light for
$\operatorname{Re}\sigma_{+}(\omega)/\sigma_{o}$ v.s. $\omega.$ The right
handed circularly polarized light in normal insulator phase only gives
$m\rightarrow\left\vert m\right\vert +1$ transition for $\tau_{z}=-1$ and
$m\rightarrow\left\vert m\right\vert -1$ transition for $\tau_{z}=+1$. In both
cases $m\leq-1$ if $\mu=0$ while for topological insulator phase it gives
$m\rightarrow\left\vert m\right\vert +1$ transition for $\tau_{z}=-1$ with
$m\leq0$ for $\mu=0$ and for $\tau_{z}=+1$ it gives $m\rightarrow\left\vert
m\right\vert -1$ transition with $m\leq-1$. The plot for topological insulator
phase is shown in Fig. 12(a).

Fig. 13 shows absorptive peaks resulted from the left handed circularly
polarized light for $\operatorname{Re}\sigma_{-}(\omega)/\sigma_{o}$ v.s.
$\omega.$ Left handed circularly polarized light gives $m\rightarrow\left\vert
m\right\vert +1$ for $\tau_{z}=+1$ and $m\rightarrow\left\vert m\right\vert
-1$ for $\tau_{z}=-1$ transitions with $m\leq0$ for $\mu=0$ in both cases.
While in topological insulator it gives $m\rightarrow\left\vert m\right\vert
+1$ for $\tau_{z}=+1$ with $m\leq0$ and $m\rightarrow\left\vert m\right\vert
-1$ for $\tau_{z}=-1$ transition with $m\leq-1$ for $\mu=0$.

\section{Semiclassical Limit}

\ The semiclassical limit is valid when the quantization between Landau levels
is unimportant. It is the case when chemical potential $\mu>>E_{1}$\cite{12}.
For large $\mu$ all transitions will be intraband. The energy of intraband
transitions is given by $\delta E=E_{n+1}-E_{n},$ which is approximated to
give%
\begin{equation}
\delta E=\frac{\hbar v_{f}^{2}eB}{\sqrt{2N\hbar v_{f}^{2}eB+(\Delta_{Z}%
\pm\Delta_{H})}}%
\end{equation}
The chemical potential $\mu$ falls exactly between $N$ and $N+1$ with $N>>1,$
so we can write $\mu\approx E_{N},$ we obtain%
\begin{equation}
\delta E=\frac{\hbar v_{f}^{2}eB}{\mu},
\end{equation}

so for $n$ to $n+1$ transitions the Re$\sigma_{xx}(\omega)$ in semiclassical
limit is written as%
\[
\frac{\operatorname{Re}\sigma_{xx}(\omega)}{\sigma_{o}}=\frac{\mu}{2\pi
}\underset{\tau_{z}=\pm1,m}{\sum}(A_{\left\vert m\right\vert +1}B_{m}%
)^{2}\times\frac{\eta}{[(\hbar\omega-\frac{\hbar v_{f}^{2}eB}{\mu})^{2}%
+\eta^{2}]}.
\]

The real part of the frequency dependent longitudinal optical conductivity
Re$\sigma_{xx}(\omega)$ versus $\hbar\omega$ in units of $e^{2}/h$ in the
semiclassical limit is shown in Fig. 14 with $E_{n}<\mu<E_{n+1}$. The first
pair starting from right side represents the transition between $n=5$ to $n=6$
while the 2nd pair represents the transition between $n=11$ and $n=12$ LLs and
the last pair at the lowest energy represents the transition between $n=20$
and $n=21$ LLs. The dashed peak gives the transition between $\tau_{z}=-1$ LLs
and solid peaks represent transitions for $\tau_{z}=+1$ LLs. We observe that
as the chemical potential increases the spectral weight increases.

\section{Broken Inversion Symmetric TI Thin film}

Thin film TIs are usually grown on a substrate which breaks inversion
symmetry. In this case, the effective Hamiltonian in the symmetric and
antisymmetric basis, Eq. (1), will be augmented by a term $V\sigma_{x}$ that
breaks inversion symmetry. $V$ represents the magnitude of inversion
asymmetry. The Hamiltonian becomes
\begin{equation}
\hat{H}_{\sigma\tau}=\sqrt{\frac{\hbar}{2}}\frac{v_{f}}{l_{B}}(i\sigma
_{x}(a^{\dagger}-a)+\tau_{z}\sigma_{y}(a+a^{\dagger}))+(\Delta_{Z}\tau
_{z}+\Delta_{H})\sigma_{z}+V\sigma_{x}.
\end{equation}
This is the inversion symmetry broken thin TI Hamiltonian given in Eq. (3) of
\cite{11}, without the exchange field for ferromagnetic ordering but including
an applied magnetic field. The single-particle eigenstates of the above
Hamiltonian have the following form:%
\begin{align}
|\left\vert n\right\vert \tau_{z}sgn(n)\rangle &  =u_{n\tau_{z}=+1}%
^{sgn(n)}|\left\vert n\right\vert -1,\uparrow,\tau_{z}=+1\rangle+u_{n\tau
_{z}=+1}^{sgn(n)}|\left\vert n\right\vert ,\downarrow,\tau_{z}=+1\rangle\\
+u_{n\tau_{z}=-1}^{sgn(n)}|\left\vert n\right\vert ,  &  \uparrow,\tau
_{z}=-1\rangle+u_{n\tau_{z}=-1}^{sgn(n)}|\left\vert n\right\vert
-1,\downarrow,\tau_{z}=-1\rangle.\nonumber
\end{align}
\textbf{ }$u_{n\tau_{z}}^{sgn(n)}$ are the complex four-component spinor wave
functions. The Landau level spectrum can be obtained by diagonalizing the
following Hamiltonian%
\begin{equation}
\left(
\begin{array}
[c]{cccc}%
\Delta_{Z}+\Delta_{H} & -i\sqrt{2\hbar n}\frac{v_{f}}{l_{B}}a & 0 & V\\
i\sqrt{2\hbar n}\frac{v_{f}}{l_{B}} & (\Delta_{Z}+\Delta_{H}) & V & 0\\
0 & V & -(\Delta_{Z}-\Delta_{H}) & i\sqrt{2\hbar n}\frac{v_{f}}{l_{B}%
}a^{\dagger}\\
V & 0 & -i\sqrt{2\hbar n}\frac{v_{f}}{l_{B}}a & \Delta_{Z}-\Delta_{H}%
\end{array}
\right)  \label{y}%
\end{equation}
Diagonalizing Eq. (\ref{y}), we find the following LL spectrum:%
\begin{equation}
E_{n}^{\tau_{z}}=sgn(n)\sqrt{2\hbar v_{f}^{2}eB\left\vert n\right\vert
+V^{2}+\Delta_{Z}^{2}+\Delta_{H}{}^{2}+\tau_{z}2\sqrt{2V^{2}\hbar v_{f}%
^{2}eB\left\vert n\right\vert +V^{2}\Delta_{Z}^{2}+\Delta_{Z}^{2}\Delta_{H}%
{}^{2}}}%
\end{equation}%
\begin{equation}
E_{0}^{\tau_{z}}=-(\Delta_{Z}+\tau_{z}\sqrt{\Delta_{H}^{2}+V^{2}}).
\end{equation}
In the inversion symmetry broken system, the phase transition from normal
insulator to topological insulating phase now occurs at $\Delta_{Z}%
=\sqrt{\Delta_{H}^{2}+V^{2}}$. This shows that the phase transition is pushed
to higher magnetic fields$.$ For $V=0$ the system is decoupled for $\tau
_{z}=\pm1.$ As a result of inversion symmetry breaking, symmetric and
antisymmetric hybridized states become coupled to each other and there can be
allowed transitions between them. An additional feature is that broken
inversion symmetry gives rise to crossing of LLs $n$ with $\tau_{z}=+1$ and
$n+1$ with $\tau_{z}=-1$ at certain values of magnetic field$,$ as shown in
Fig. 15. This will allow additional transitions between LLs with $\tau_{z}=+1$
and LLs with $\tau_{z}=-1.$ Recently, one photon and two photon absorption was
investigated in topological insulator thin films with broken inversion
symmetry \cite{41}, in the absence of a magnetic field. It was observed that
additional transition channels open when inversion symmetry is not present.

\section{\textbf{\bigskip Effect of Hexagonal Warping on Landau Levels and
Magneto-Optical Conductivity}}

ARPES data suggests that there is warping effect on the band structure of
$Bi_{2}Te_{3}$. To take this effect into account, Fu\cite{42} added a cubic
correction term in the Hamiltonian of a topological insulator. This induces an
anisotropic effect on the bands in momentum space whose strength is set by the
parameter $\lambda$. The Hamiltonian of TI thin film in a magnetic field with
a warping term is%
\begin{equation}
\hat{H}_{\sigma\tau}^{\prime}=\hat{H}_{\sigma\tau}-\frac{2\lambda\hbar^{3/2}%
}{l_{B}^{3}}[(a^{\dagger})^{3}+(a)^{3}]\sigma_{z}\tau_{z},
\end{equation}
where $\hat{H}_{\sigma\tau}$ is given in Eq. (\ref{x}). Analytical
diagonalization of the above Hamiltonian is not possible. However, if warping
is treated as a perturbation, then effect of warping on Landau levels can be
evaluated\cite{43}. These results show that the primary effect of warping on
the LL spectrum is that slope of LLs increases with increasing magnetic field.
This has implications on our work; warping affects the gap between LLs with
optical transitions shifting to higher $\hbar\omega$. This effect will be more
significant at higher magnetic fields.

\section{Summary and Conclusions}

We have studied the effect of hybridization between the top and bottom surface
states on the magneto-optical conductivity in a thin film TI . Hybridization
induces gap in the Dirac spectrum. Each LL splits into two with the same
Landau index representing LLs for symmetric and antisymmetric hybridized
states. At a critical magnetic field the system makes a quantum phase
transition from the NI phase to a TI phase. This has a signature in the
magneto-optical absorption spectra, both $\operatorname{Re}\sigma_{xx}%
(\omega)$ and $\operatorname{Im}\sigma_{xy}(\omega)$. $\operatorname{Re}%
\sigma_{xx}(\omega)$ peak for $E_{-1}^{-}\rightarrow E_{0}^{-}$ transition in
NI phase is replaced by~$E_{0}^{-}\rightarrow E_{1}^{-}$ peak in TI phase.
More significant signature for the quantum phase transition is found in
absorption spectra for $\operatorname{Im}\sigma_{xy}(\omega)$. It shows
negative peaks in the TI phase which are absent in the absorption spectra in
NI peaks. The signature in circularly polarized light is the splitting and
shifting of the absorption peaks in TI phase relative to the NI phase.

\section{Acknowledgement}

A. Ullah and K. Sabeeh acknowledge the support of Higher Education Commission
(HEC) of Pakistan through project No. 20-1484/R\&D/09. K. Sabeeh would also
like to acknowledge the support of the Abdus Salam International Center for
Theoretical Physics (ICTP) in Trieste, Italy through the Associate Scheme
where a part of this work was completed.

%

%TCIMACRO{\FRAME{ftbpFU}{6.832in}{5.2866in}{0pt}{\Qcb{Landau level energies as
%a function of magnetic field(B) in units of Tesla with hydridization energy
%$\Delta_{H}=0.004eV$ and Zeeman energy $\Delta_{Z}=0.00174\times B\frac{eV}%
%{T}$.}}{}{1.eps}{\special{ language "Scientific Word";  type "GRAPHIC";
%width 6.832in;  height 5.2866in;  depth 0pt;  original-width 11.0056in;
%original-height 8.4968in;  cropleft "0";  croptop "1";  cropright "1";
%cropbottom "0";  filename '1.eps';file-properties "XNPEU";}}}%
%BeginExpansion
\begin{figure}
[ptb]
\begin{center}
\includegraphics[
height=5.2866in,
width=6.832in
]%
{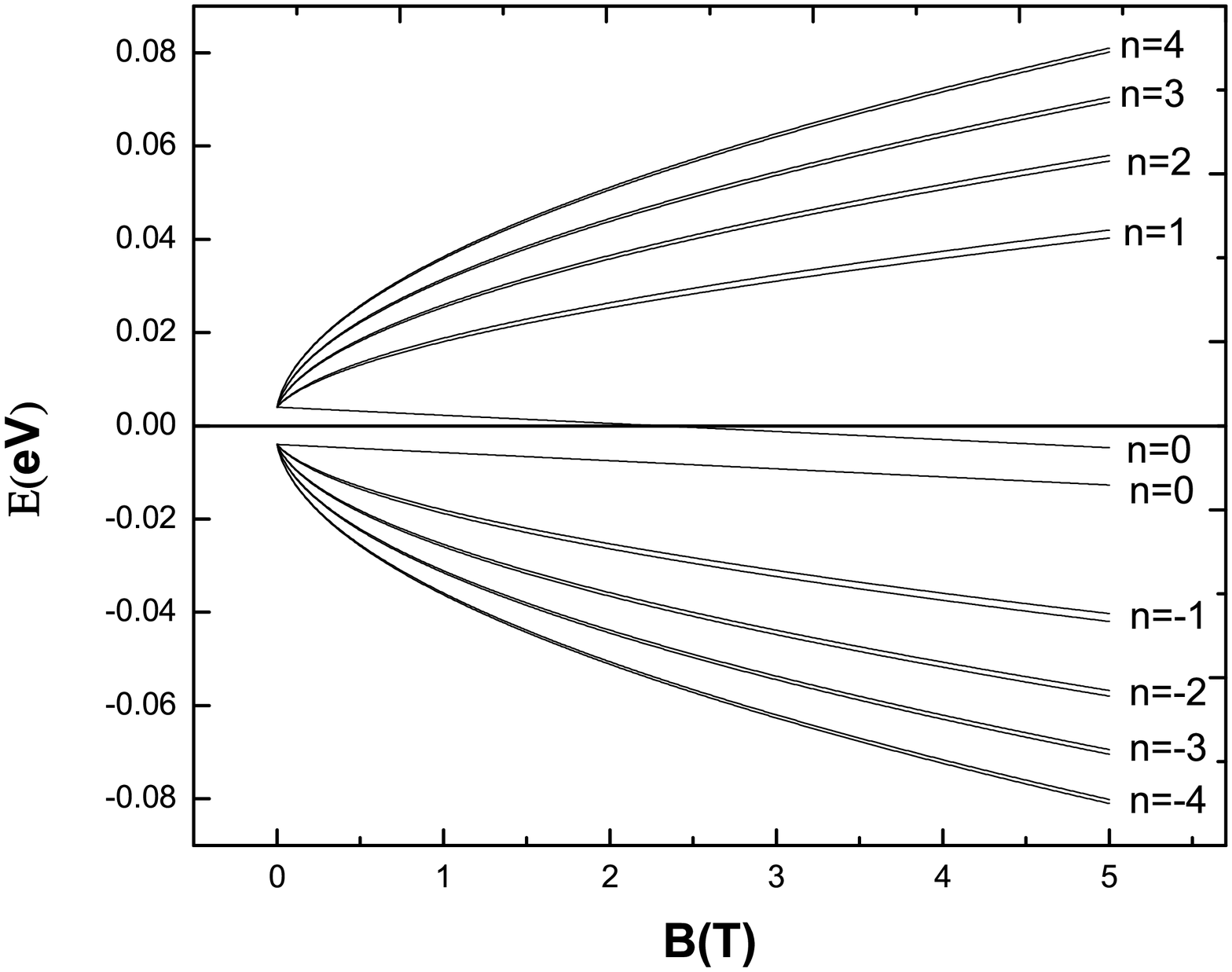}%
\caption{Landau level energies as a function of magnetic field(B) in units of
Tesla with hydridization energy $\Delta_{H}=0.004eV$ and Zeeman energy
$\Delta_{Z}=0.00174\times B\frac{eV}{T}$.}%
\end{center}
\end{figure}
%EndExpansion
%TCIMACRO{\FRAME{ftbpFU}{5.2607in}{6.7931in}{0pt}{\Qcb{Density of states for
%thin film topological insulator in a magnetic field in units of eB/2$\pi\hbar
%$. (a) Density of states in topological insulator phase $(\Delta_{Z}%
%<\Delta_{H})$ (b) Density of states at charge neutrality point $(\Delta
%_{Z}=\Delta_{H})$. (c) Density of states in normal insulator phase
%$(\Delta_{Z}>\Delta_{H})$.}}{}{2.eps}{\special{ language "Scientific Word";
%type "GRAPHIC";  maintain-aspect-ratio TRUE;  display "USEDEF";
%valid_file "F";  width 5.2607in;  height 6.7931in;  depth 0pt;
%original-width 8.4968in;  original-height 11.0056in;  cropleft "0";
%filename '2.eps';file-properties "XNPEU";}}}%
%BeginExpansion
\begin{figure}
[ptb]
\begin{center}
\includegraphics[
height=6.7931in,
width=5.2607in
]%
{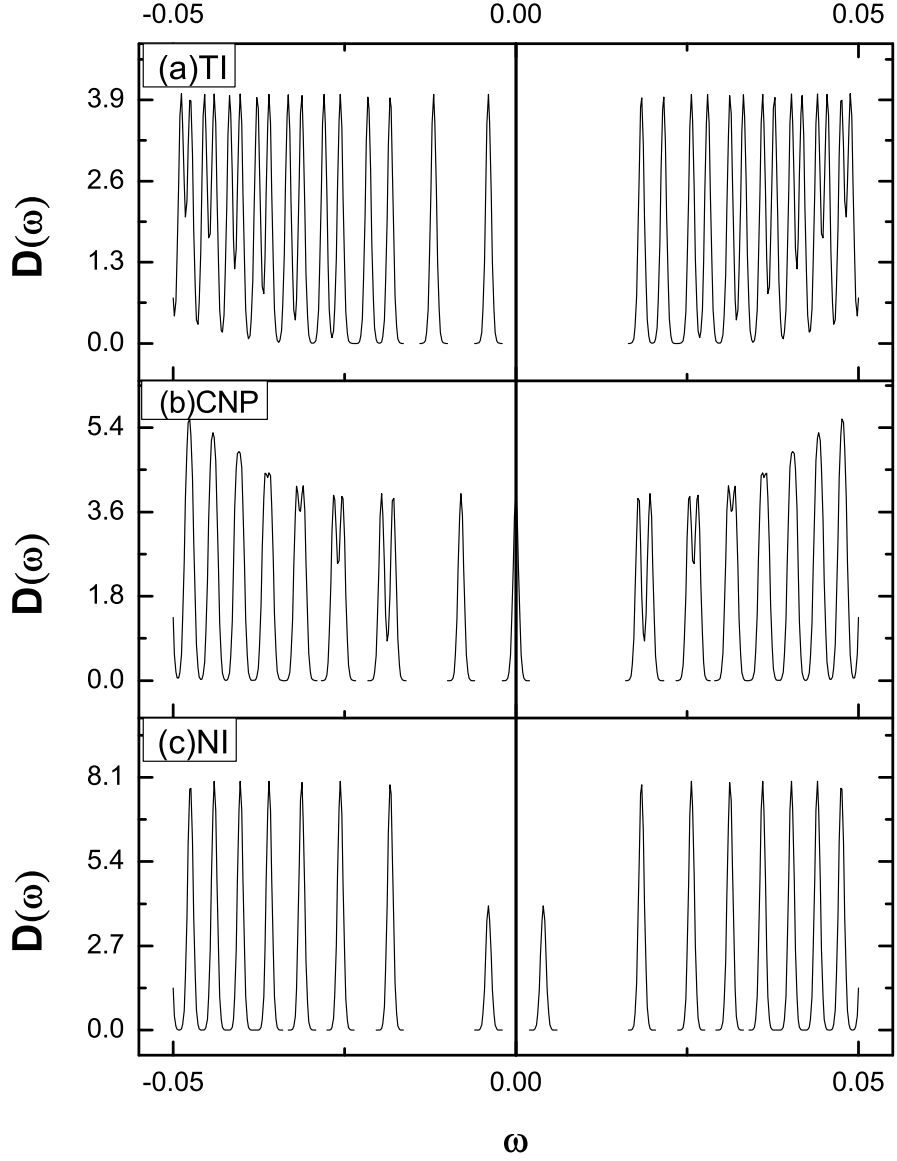}%
\caption{Density of states for thin film topological insulator in a magnetic
field in units of eB/2$\pi\hbar$. (a) Density of states in topological
insulator phase $(\Delta_{Z}<\Delta_{H})$ (b) Density of states at charge
neutrality point $(\Delta_{Z}=\Delta_{H})$. (c) Density of states in normal
insulator phase $(\Delta_{Z}>\Delta_{H})$.}%
\end{center}
\end{figure}
%EndExpansion
%TCIMACRO{\FRAME{ftbpFU}{5.1664in}{6.6729in}{0pt}{\Qcb{Real part of the
%longitudinal conductivity $\sigma_{xx}(\omega)$ of thin film topological
%insulator in units of $e^{2}/\hbar$ as a function of $\hbar\omega$ in $eV$
%compared in (a) topological insulator phase, (b) CNP and (c) normal insulator
%phase. The scattering rate is $\eta=0.15\Delta_{H}$ and $\mu=0$.}}{}%
%{3.eps}{\special{ language "Scientific Word";  type "GRAPHIC";
%width 5.1664in;  height 6.6729in;  depth 0pt;  original-width 8.4968in;
%original-height 11.0056in;  cropleft "0";  croptop "1";  cropright "1";
%cropbottom "0";  filename '3.eps';file-properties "XNPEU";}}}%
%BeginExpansion
\begin{figure}
[ptb]
\begin{center}
\includegraphics[
height=6.6729in,
width=5.1664in
]%
{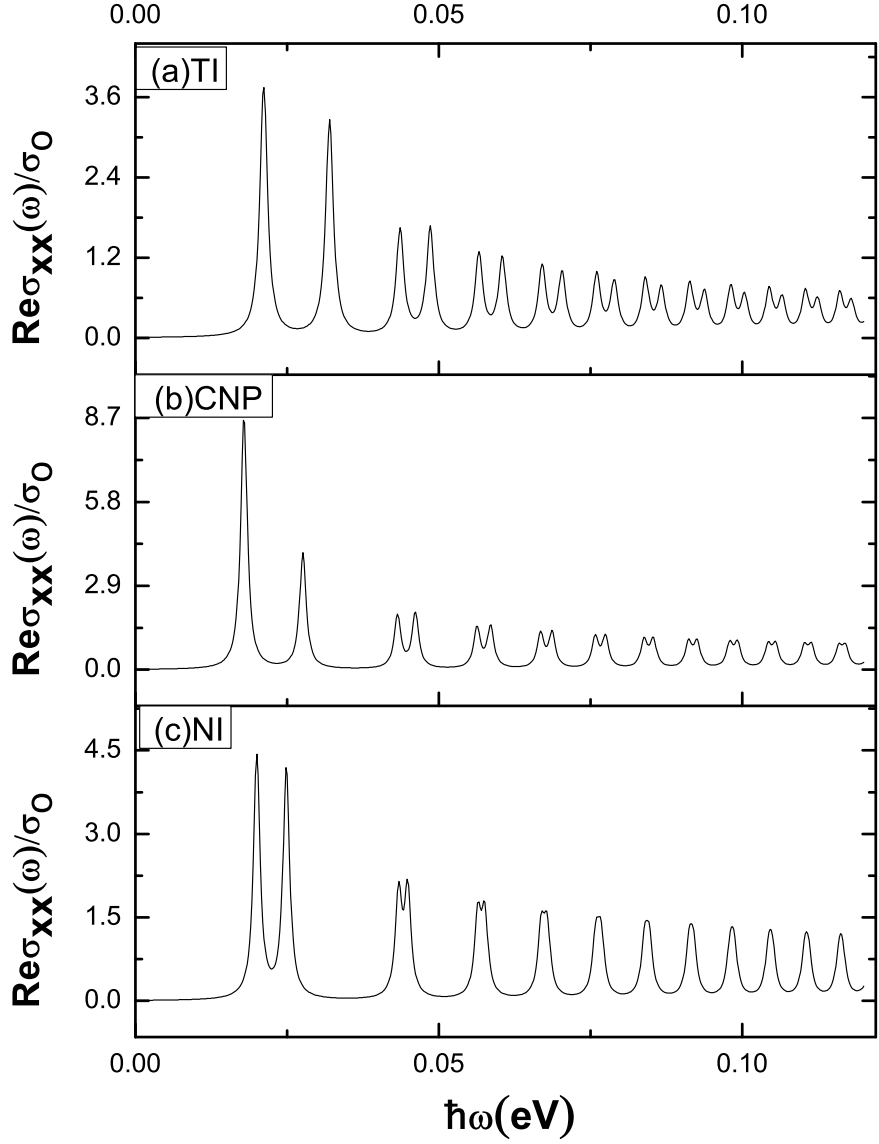}%
\caption{Real part of the longitudinal conductivity $\sigma_{xx}(\omega)$ of
thin film topological insulator in units of $e^{2}/\hbar$ as a function of
$\hbar\omega$ in $eV$ compared in (a) topological insulator phase, (b) CNP and
(c) normal insulator phase. The scattering rate is $\eta=0.15\Delta_{H}$ and
$\mu=0$.}%
\end{center}
\end{figure}
%EndExpansion
%TCIMACRO{\FRAME{ftbpFU}{5.1041in}{6.5916in}{0pt}{\Qcb{Imaginary part of the
%transverse conductivity $\sigma_{xy}(\omega)$ of thin film topological
%insulator in units of $e^{2}/\hbar$ as a function of $\hbar\omega$ in $eV$
%compared in (a) topological insulator phase, (b) CNP and (c) normal insulator
%phase. The scattering rate is $\eta=0.15\Delta_{H}$ and $\mu=0$.}}{}%
%{4.eps}{\special{ language "Scientific Word";  type "GRAPHIC";
%width 5.1041in;  height 6.5916in;  depth 0pt;  original-width 8.4968in;
%original-height 11.0056in;  cropleft "0";  croptop "1";  cropright "1";
%cropbottom "0";  filename '4.eps';file-properties "XNPEU";}}}%
%BeginExpansion
\begin{figure}
[ptb]
\begin{center}
\includegraphics[
height=6.5916in,
width=5.1041in
]%
{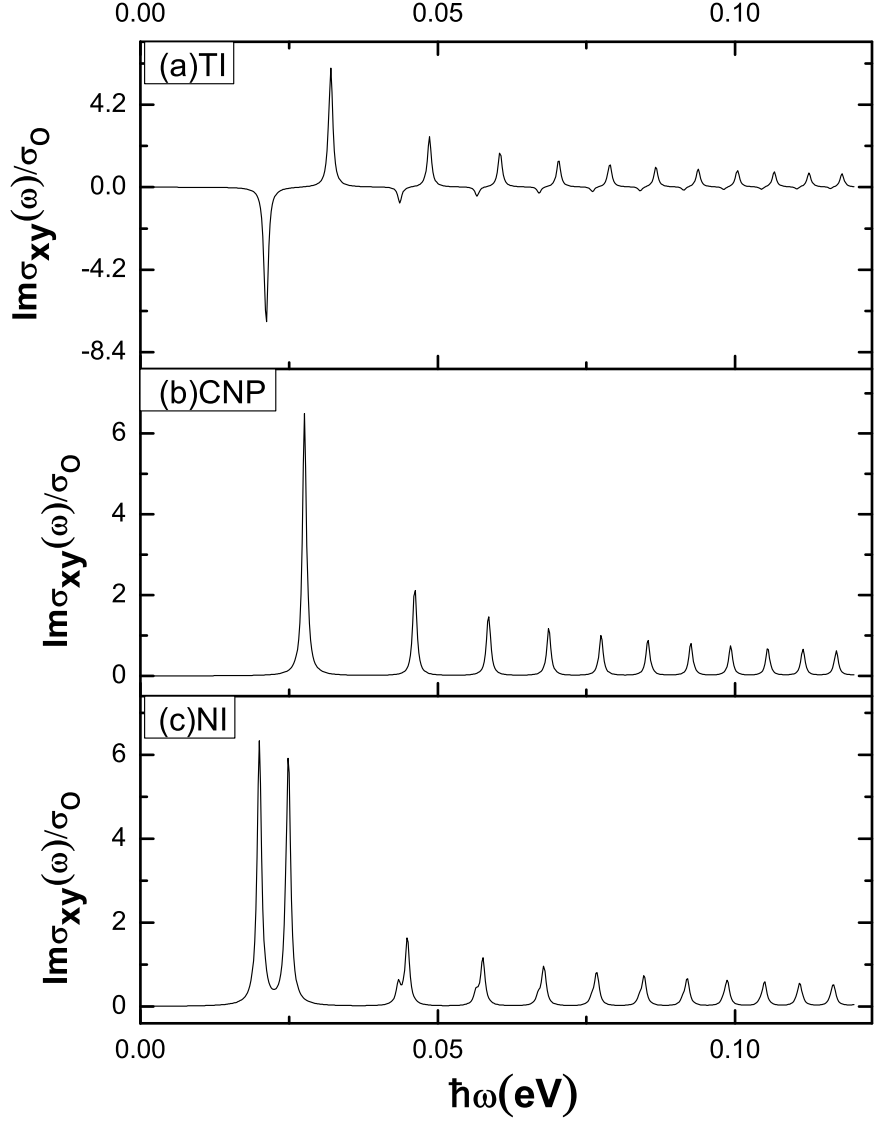}%
\caption{Imaginary part of the transverse conductivity $\sigma_{xy}(\omega)$
of thin film topological insulator in units of $e^{2}/\hbar$ as a function of
$\hbar\omega$ in $eV$ compared in (a) topological insulator phase, (b) CNP and
(c) normal insulator phase. The scattering rate is $\eta=0.15\Delta_{H}$ and
$\mu=0$.}%
\end{center}
\end{figure}
%EndExpansion
%TCIMACRO{\FRAME{ftbpFU}{6.2578in}{7.5809in}{0pt}{\Qcb{Schematic representation
%of the allowed transitions between Landau level of same $\tau_{z}$ in normal
%insulator phase for $\mu=0$. Blue lines represent Landau levels for $\tau
%_{z}=-1$ and red lines represent Landau levels for $\tau_{z}=+1$.}}{}%
%{5.eps}{\special{ language "Scientific Word";  type "GRAPHIC";
%width 6.2578in;  height 7.5809in;  depth 0pt;  original-width 3.3338in;
%original-height 4.0473in;  cropleft "0";  croptop "1";  cropright "1";
%cropbottom "0";  filename '5.eps';file-properties "XNPEU";}}}%
%BeginExpansion
\begin{figure}
[ptb]
\begin{center}
\includegraphics[
height=7.5809in,
width=6.2578in
]%
{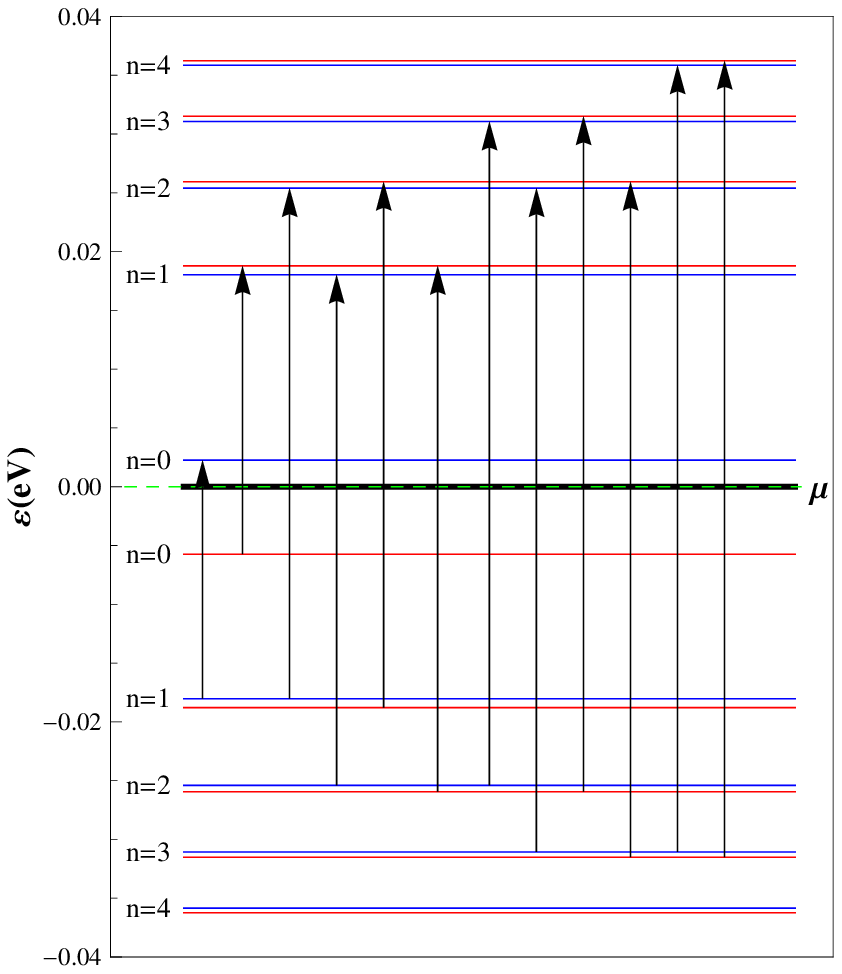}%
\caption{Schematic representation of the allowed transitions between Landau
level of same $\tau_{z}$ in normal insulator phase for $\mu=0$. Blue lines
represent Landau levels for $\tau_{z}=-1$ and red lines represent Landau
levels for $\tau_{z}=+1$.}%
\end{center}
\end{figure}
%EndExpansion
%TCIMACRO{\FRAME{ftbpFU}{3.7905in}{5.9612in}{0pt}{\Qcb{Schematic representation
%of the allowed transitions between Landau level of same $\tau_{z}$ at CNP
%$\mu=0$. Blue lines represent Landau levels for $\tau_{z}=-1$ and red lines
%represent Landau levels for $\tau_{z}=+1$.}}{}{6.eps}%
%width 3.7905in;  height 5.9612in;  depth 0pt;  original-width 3.3338in;
%original-height 5.2762in;  cropleft "0";  croptop "1";  cropright "1";
%cropbottom "0";  filename '6.eps';file-properties "XNPEU";}}}%
%BeginExpansion
\begin{figure}
[ptb]
\begin{center}
\includegraphics[
height=5.9612in,
width=3.7905in
]%
{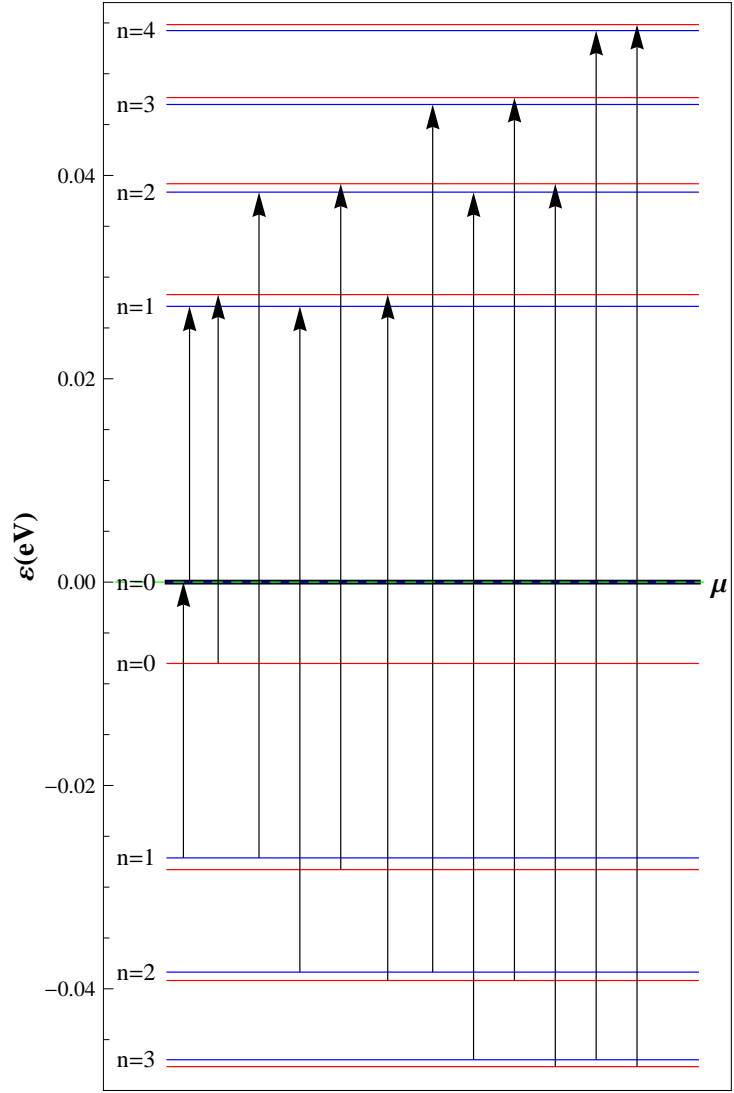}%
\caption{Schematic representation of the allowed transitions between Landau
level of same $\tau_{z}$ at CNP $\mu=0$. Blue lines represent Landau levels
for $\tau_{z}=-1$ and red lines represent Landau levels for $\tau_{z}=+1$.}%
\end{center}
\end{figure}
%EndExpansion
%TCIMACRO{\FRAME{ftbpFU}{4.0689in}{7.2437in}{0pt}{\Qcb{Schematic representation
%of the allowed transitions between Landau level of same $\tau_{z}$ in
%topological insulator phase. Blue lines represent Landau levels for $\tau
%_{z}=-1$ and red lines represent Landau levels for $\tau_{z}=+1$.}}{}%
%{7.eps}{\special{ language "Scientific Word";  type "GRAPHIC";
%width 4.0689in;  height 7.2437in;  depth 0pt;  original-width 3.3338in;
%original-height 5.988in;  cropleft "0";  croptop "1";  cropright "1";
%cropbottom "0";  filename '7.eps';file-properties "XNPEU";}}}%
%BeginExpansion
\begin{figure}
[ptb]
\begin{center}
\includegraphics[
height=7.2437in,
width=4.0689in
]%
{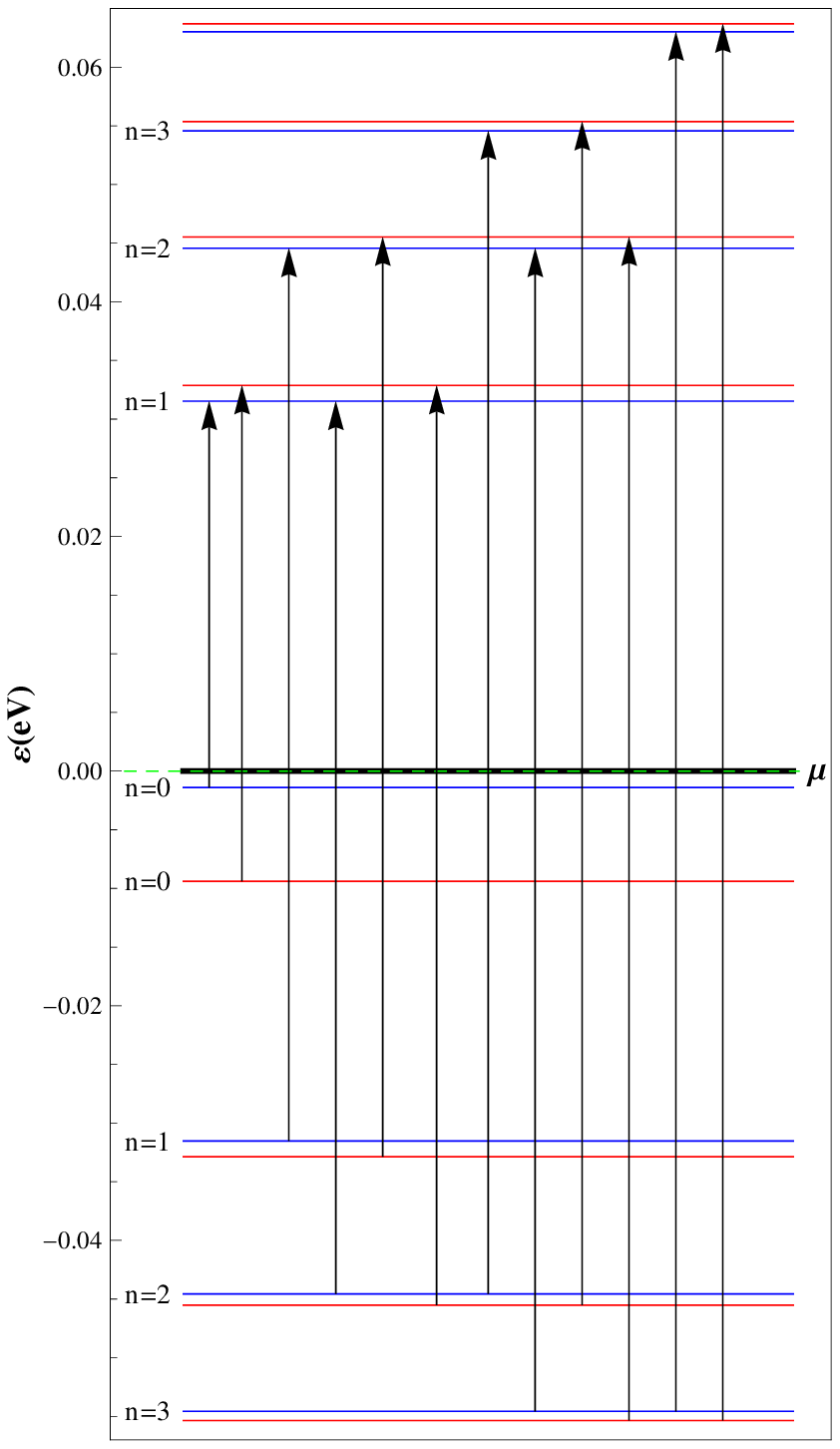}%
\caption{Schematic representation of the allowed transitions between Landau
level of same $\tau_{z}$ in topological insulator phase. Blue lines represent
Landau levels for $\tau_{z}=-1$ and red lines represent Landau levels for
$\tau_{z}=+1$.}%
\end{center}
\end{figure}
%EndExpansion
%TCIMACRO{\FRAME{ftbpFU}{5.3722in}{4.158in}{0pt}{\Qcb{Real part of the
%longitudinal conductivity $\sigma_{xx}(\omega)$ of thin film topological
%insulator in units of $e^{2}/\hbar$ as a function of $\hbar\omega$ in $eV$ in
%normal insulator phase. The scattering rate is $\eta=0.15\Delta_{H}$ and
%$\mu=0.0225$ and $B=2T.$ Red peak represents absorption peak for intraband
%transition while black peaks represent absorption peaks for interband
%transitions.}}{}{8.eps}{\special{ language "Scientific Word";
%type "GRAPHIC";  maintain-aspect-ratio TRUE;  display "USEDEF";
%valid_file "F";  width 5.3722in;  height 4.158in;  depth 0pt;
%original-width 11.0056in;  original-height 8.4968in;  cropleft "0";
%filename '8.eps';file-properties "XNPEU";}}}%
%BeginExpansion
\begin{figure}
[ptb]
\begin{center}
\includegraphics[
height=4.158in,
width=5.3722in
]%
{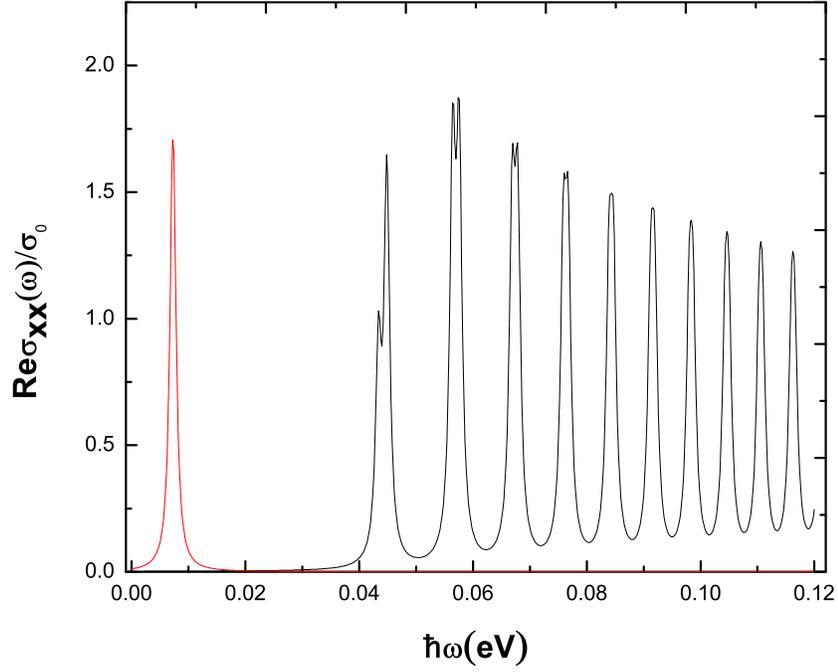}%
\caption{Real part of the longitudinal conductivity $\sigma_{xx}(\omega)$ of
thin film topological insulator in units of $e^{2}/\hbar$ as a function of
$\hbar\omega$ in $eV$ in normal insulator phase. The scattering rate is
$\eta=0.15\Delta_{H}$ and $\mu=0.0225$ and $B=2T.$ Red peak represents
absorption peak for intraband transition while black peaks represent
absorption peaks for interband transitions.}%
\end{center}
\end{figure}
%EndExpansion
%TCIMACRO{\FRAME{ftbpFU}{5.2684in}{4.0776in}{0pt}{\Qcb{Real part of the
%longitudinal conductivity $\sigma_{xx}(\omega)$ of thin film topological
%insulator in units of $e^{2}/\hbar$ as a function of $\hbar\omega$ in $eV$ in
%topological insulator phase. The scattering rate is $\eta=0.15\Delta_{H}$,
%$\mu=0.0225$ and $B=4T.$ All peaks represent absorption peaks for interband
%transitions.}}{}{9.eps}{\special{ language "Scientific Word";
%type "GRAPHIC";  maintain-aspect-ratio TRUE;  display "USEDEF";
%valid_file "F";  width 5.2684in;  height 4.0776in;  depth 0pt;
%original-width 11.0056in;  original-height 8.4968in;  cropleft "0";
%filename '9.eps';file-properties "XNPEU";}}}%
%BeginExpansion
\begin{figure}
[ptb]
\begin{center}
\includegraphics[
height=4.0776in,
width=5.2684in
]%
{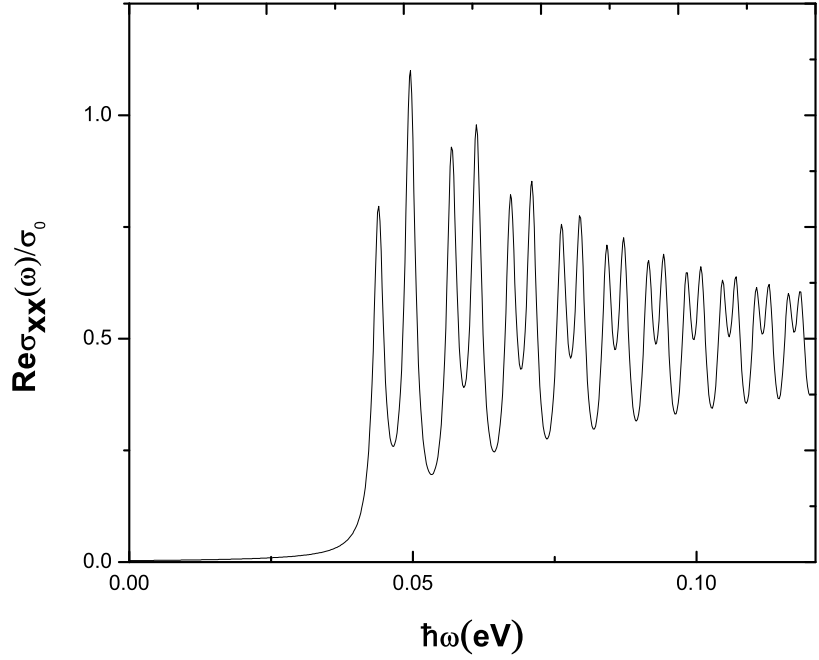}%
\caption{Real part of the longitudinal conductivity $\sigma_{xx}(\omega)$ of
thin film topological insulator in units of $e^{2}/\hbar$ as a function of
$\hbar\omega$ in $eV$ in topological insulator phase. The scattering rate is
$\eta=0.15\Delta_{H}$, $\mu=0.0225$ and $B=4T.$ All peaks represent absorption
peaks for interband transitions.}%
\end{center}
\end{figure}
%EndExpansion
%TCIMACRO{\FRAME{ftbpFU}{3.9522in}{4.0378in}{0pt}{\Qcb{Schematic representation
%of the allowed transitions between Landau levels of same $\tau_{z}$ in normal
%insulator phase for $B=1T$ and $\mu=0.02eV$. Orange arrows represent intraband
%transition and black arrows represent interband transitions.}}{}%
%{10.eps}{\special{ language "Scientific Word";  type "GRAPHIC";
%width 3.9522in;  height 4.0378in;  depth 0pt;  original-width 3.3304in;
%original-height 3.4039in;  cropleft "0";  croptop "1";  cropright "1";
%cropbottom "0";  filename '10.eps';file-properties "XNPEU";}}}%
%BeginExpansion
\begin{figure}
[ptb]
\begin{center}
\includegraphics[
height=4.0378in,
width=3.9522in
]%
{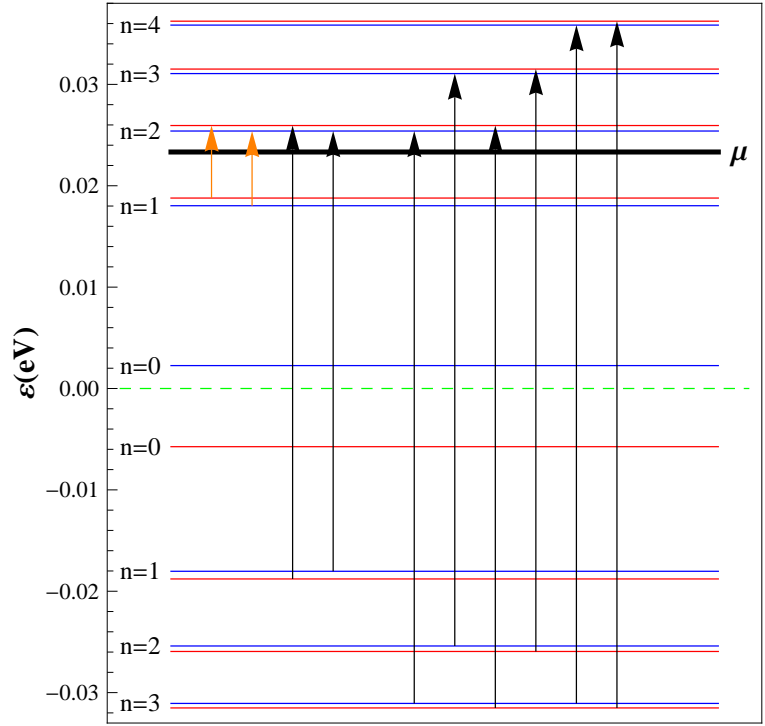}%
\caption{Schematic representation of the allowed transitions between Landau
levels of same $\tau_{z}$ in normal insulator phase for $B=1T$ and
$\mu=0.02eV$. Orange arrows represent intraband transition and black arrows
represent interband transitions.}%
\end{center}
\end{figure}
%EndExpansion
%TCIMACRO{\FRAME{ftbpFU}{3.8026in}{6.6028in}{0pt}{\Qcb{Schematic representation
%of the allowed transitions between Landau level of same $\tau_{z}$ in
%topological insulator phase with $B=3T$ and $\mu=0.02eV$.}}{}{11.eps}%
%width 3.8026in;  height 6.6028in;  depth 0pt;  original-width 3.3338in;
%original-height 5.834in;  cropleft "0";  croptop "1";  cropright "1";
%cropbottom "0";  filename '11.eps';file-properties "XNPEU";}}}%
%BeginExpansion
\begin{figure}
[ptb]
\begin{center}
\includegraphics[
height=6.6028in,
width=3.8026in
]%
{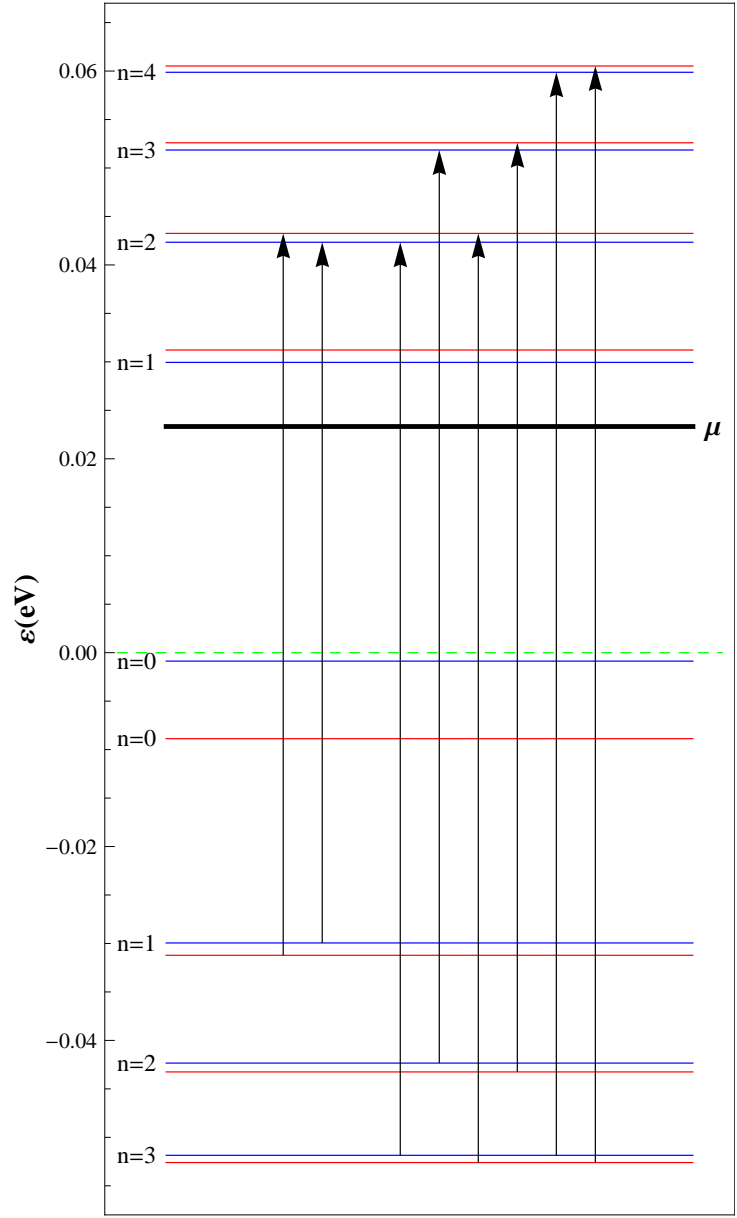}%
\caption{Schematic representation of the allowed transitions between Landau
level of same $\tau_{z}$ in topological insulator phase with $B=3T$ and
$\mu=0.02eV$.}%
\end{center}
\end{figure}
%EndExpansion
%TCIMACRO{\FRAME{ftbpFU}{4.3751in}{5.649in}{0pt}{\Qcb{$Re\sigma_{+}%
%(\omega)/\sigma_{o}$ as a function of frequency for right handed
%circularly-polarized light in (a) topological insulator phase (b) at CNP (c)
%normal insulator phase.}}{}{12.eps}{\special{ language "Scientific Word";
%type "GRAPHIC";  maintain-aspect-ratio TRUE;  display "USEDEF";
%valid_file "F";  width 4.3751in;  height 5.649in;  depth 0pt;
%original-width 8.4968in;  original-height 11.0056in;  cropleft "0";
%filename '12.eps';file-properties "XNPEU";}}}%
%BeginExpansion
\begin{figure}
[ptb]
\begin{center}
\includegraphics[
height=5.649in,
width=4.3751in
]%
{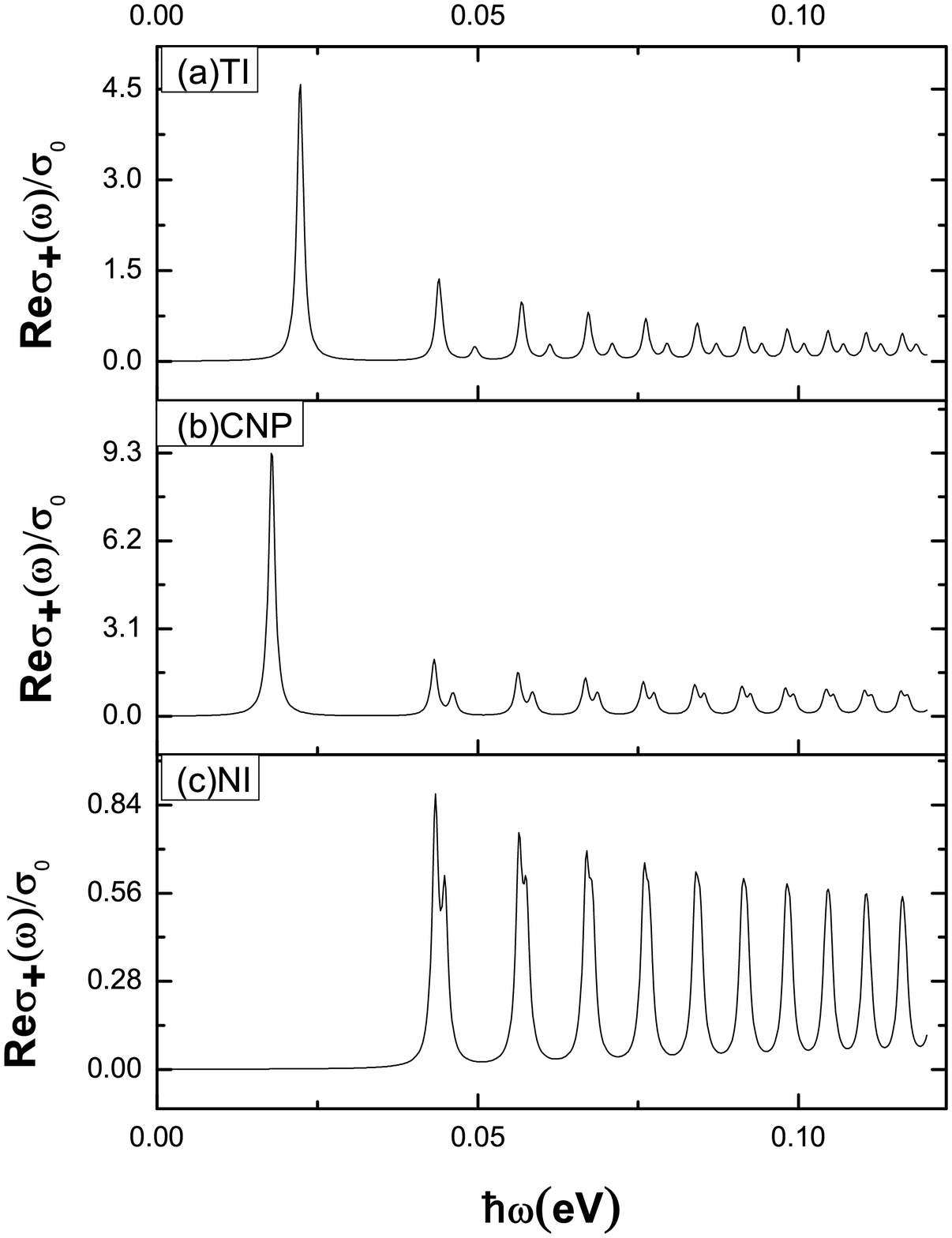}%
\caption{$Re\sigma_{+}(\omega)/\sigma_{o}$ as a function of frequency for
right handed circularly-polarized light in (a) topological insulator phase (b)
at CNP (c) normal insulator phase.}%
\end{center}
\end{figure}
%EndExpansion
%TCIMACRO{\FRAME{ftbpFU}{4.6873in}{6.0528in}{0pt}{\Qcb{$Re\sigma_{-}%
%(\omega)/\sigma_{o}$ as a function of frequency for left handed
%circularly-polarized light in (a) topological insulator phase (b) at CNP (c)
%normal insulator phase.}}{}{13.eps}{\special{ language "Scientific Word";
%type "GRAPHIC";  maintain-aspect-ratio TRUE;  display "USEDEF";
%valid_file "F";  width 4.6873in;  height 6.0528in;  depth 0pt;
%original-width 8.4968in;  original-height 11.0056in;  cropleft "0";
%filename '13.eps';file-properties "XNPEU";}}}%
%BeginExpansion
\begin{figure}
[ptb]
\begin{center}
\includegraphics[
height=6.0528in,
width=4.6873in
]%
{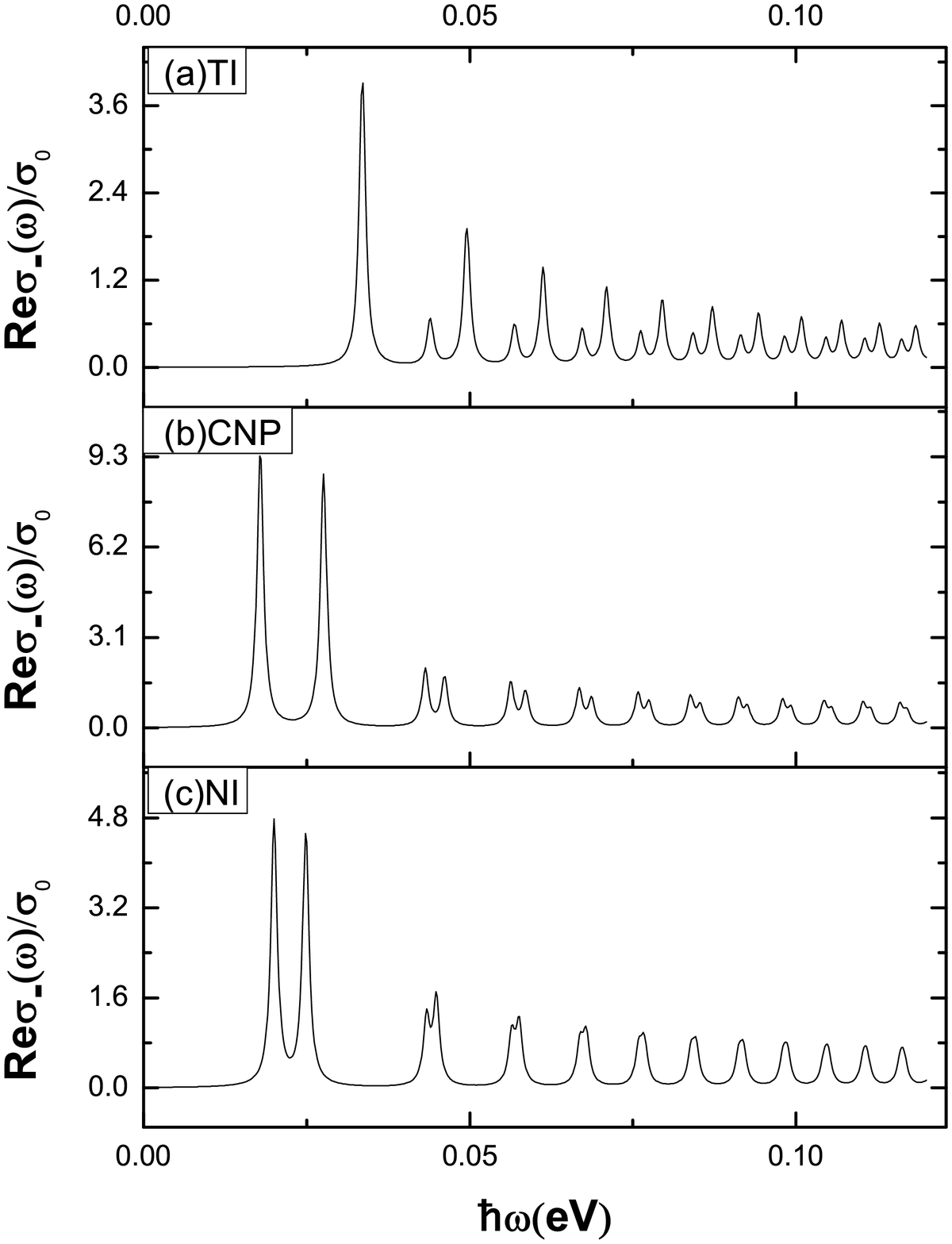}%
\caption{$Re\sigma_{-}(\omega)/\sigma_{o}$ as a function of frequency for left
handed circularly-polarized light in (a) topological insulator phase (b) at
CNP (c) normal insulator phase.}%
\end{center}
\end{figure}
%EndExpansion
%TCIMACRO{\FRAME{ftbpFU}{5.0938in}{3.3338in}{0pt}{\Qcb{The semiclassical limit
%of the real part of the longitudinal conductivity $\operatorname{Re}%
%\sigma_{xx}(\omega)/\sigma_{o}$ in units of $e^{2}/\hbar$ in eV.}}{}%
%{14.eps}{\special{ language "Scientific Word";  type "GRAPHIC";
%width 5.0938in;  height 3.3338in;  depth 0pt;  original-width 6.2111in;
%original-height 4.0499in;  cropleft "0";  croptop "1";  cropright "1";
%cropbottom "0";  filename '14.eps';file-properties "XNPEU";}}}%
%BeginExpansion
\begin{figure}
[ptb]
\begin{center}
\includegraphics[
height=3.3338in,
width=5.0938in
]%
{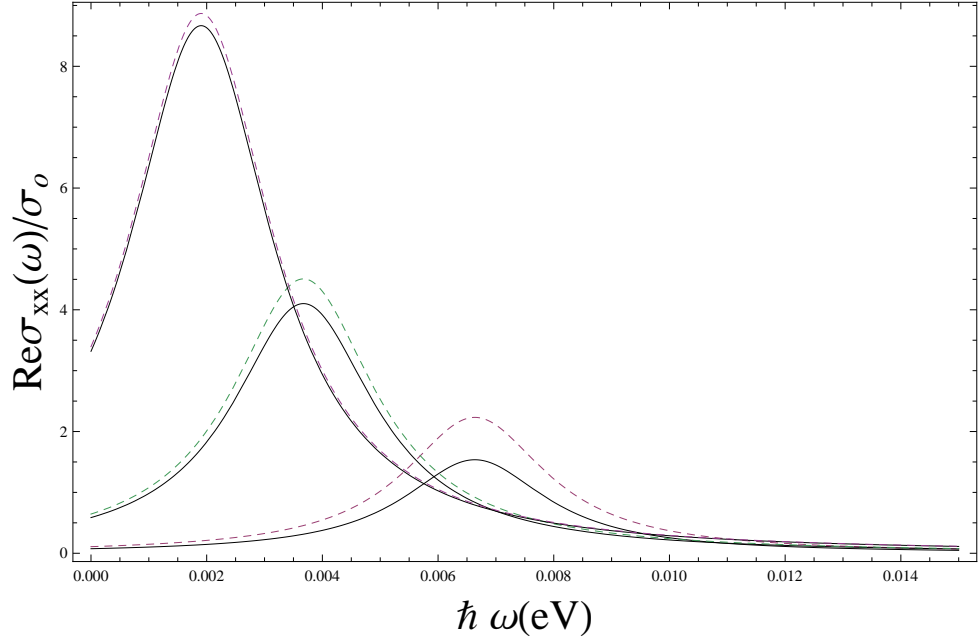}%
\caption{The semiclassical limit of the real part of the longitudinal
conductivity $\operatorname{Re}\sigma_{xx}(\omega)/\sigma_{o}$ in units of
$e^{2}/\hbar$ in eV.}%
\end{center}
\end{figure}
%EndExpansion
%TCIMACRO{\FRAME{ftbpFU}{5.5979in}{3.7853in}{0pt}{\Qcb{Landau levels energies
%for broken inversion symmetric TI as a function of magnetic field(B) in units
%of Tesla with hybridization energy $\Delta_{H}=0.004eV$, Zeeman energy
%$\Delta_{Z}=0.00174\times B\frac{eV}{T}$ and $V=0.006eV$.}}{}{15.eps}%
%width 5.5979in;  height 3.7853in;  depth 0pt;  original-width 3.333in;
%original-height 2.2459in;  cropleft "0";  croptop "1";  cropright "1";
%cropbottom "0";  filename '15.eps';file-properties "XNPEU";}}}%
%BeginExpansion
\begin{figure}
[ptb]
\begin{center}
\includegraphics[
height=3.7853in,
width=5.5979in
]%
{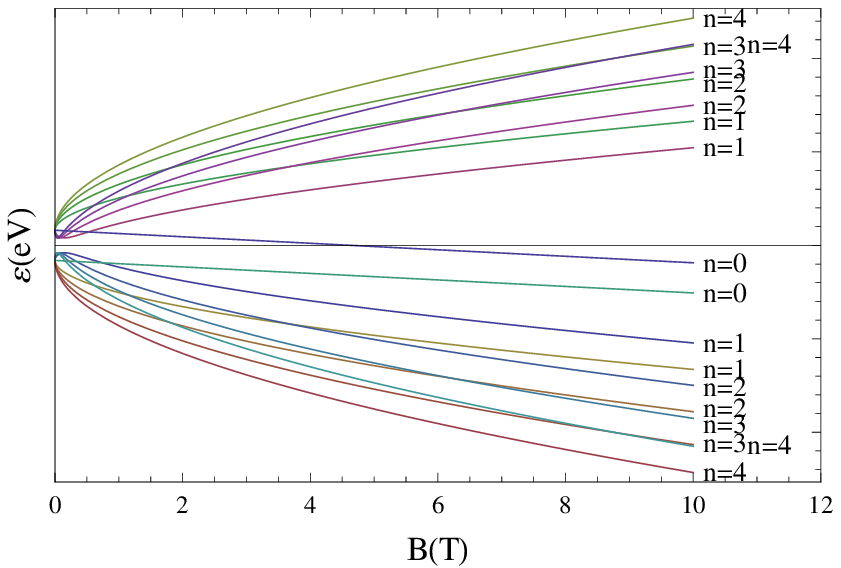}%
\caption{Landau levels energies for broken inversion symmetric TI as a
function of magnetic field(B) in units of Tesla with hybridization energy
$\Delta_{H}=0.004eV$, Zeeman energy $\Delta_{Z}=0.00174\times B\frac{eV}{T}$
and $V=0.006eV$.}%
\end{center}
\end{figure}
%EndExpansion


\begin{thebibliography}{99}                                                                                               %


\bibitem {1}M. Z. Hasan and C. L. Kane, Rev. Mod. Phys. \textbf{82}, 3045 (2010).

\bibitem {2}X.-L. Qi and S.-C. Zhang, Rev. Mod. Phys. \textbf{83}, 1057 (2011).

\bibitem {3}D. Hsieh, et. al, Nature (London) \textbf{452}, 970 (2008).

\bibitem {4}Y. L. Chen, J. G. Analytiset al., Science \textbf{325}, 178 (2009).

\bibitem {5}D. Hsieh, et. al, Nature (London) \textbf{460}, 1101 (2009).

\bibitem {6}C. Jozwiak, X. L. Chenet al.,Phys. Rev. B \textbf{84}, 165113 (2011).

\bibitem {7}S.-Y. Xu, X. Xiaet al.,Science \textbf{332}, 560 (2011).

\bibitem {8}P. Ghaemi, R. Mong, and J.E. Moore, Phys. Rev. Lett. \textbf{105},
166603 (2010).

\bibitem {9}H.-Z. Lu, W.-Y. Shan, W. Yao, Q. Niu, and S.-Q. Shen, Phys. Rev. B
\textbf{81}, 115407 (2010).

\bibitem {10}C.-X. Liu, H. Zhang, B. Yan, X.-L. Qi, T. Frauenheim, X. Dai, Z.
Fang, and S.-C. Zhang, Phys. Rev. B \textbf{81}, 041307(R) (2010).

\bibitem {11}R. Yu, W. Zhang, H.-J. Zhang, S.-C. Zhang, X. Dai, and Z.
Fang,Science \textbf{329}, 61 (2010).

\bibitem {12}B. Seradjeh, J.E. Moore, and M. Franz, Phys. Rev. Lett.
\textbf{103}, 066402 (2010).

\bibitem {13}Y. Zhang, K. He, C. Z. Chang, C. L. Song, L. L. Wang, X. Chen, J.
F. Jia, Z. Fang, X. Dai, W. Y. Shan, S. Q. Shen, Q. Niu, X. L. Qi, S. C.
Zhang, X. C. Ma, and Q. K. Xue, Nat. Phys. \textbf{6}, 584 (2010).

\bibitem {14}H. Cao, J. Tian, I. Miotkowski, T. Shen, J. Hu, S. Qiao, and Y.
P. Chen, Phys. Rev. Lett. \textbf{108}, 216803 (2012).

\bibitem {15}C. X. Liu, H. Zhang, B. Yan, X. L. Qi, T. Frauenheim, X. Dai, Z.
Fang, and S. C. Zhang, Phys. Rev. B \textbf{81}, 041307 (2010).

\bibitem {16}G. Zhang, H. Qin, J. Teng, J. Guo, Q. Guo, X. Dai, Z. Fang, and
K. Wu, Appl. Phys. Lett. \textbf{95}, 053114 (2009).

\bibitem {17}P. Cheng, C. Song, T. Zhang, Y. Zhang, Y. Wang, J. F. Jia, J.
Wang, Y. Wang, B. F. Zhu, X. Chen, X. Ma, K. He, L. Wang, X. Dai, Z. Fang, X.
Xie, X. L. Qi, C. X. Liu, S. C. Zhang, and Q. K. Xue, Phys. Rev. Lett.
\textbf{105}, 076801 (2010).

\bibitem {18}J. Wang, A. M. DaSilva, C. Z. Chang, K. He, J. K. Jain, N.
Samarth, X. C. Ma, Q. K. Xue, and M. H. W. Chan, Phys. Rev. B \textbf{83},
245438 (2011).

\bibitem {19}Jacob Linder, Takehito Yokoyama, and Asle Sudbo, Phys. Rev. B
\textbf{80}, 205401 (2009).

\bibitem {20}A. A. Zyuzin, M. D. Hook, and A. A. Burkov, Phys. Rev. B
\textbf{83}, 245428 (2011).

\bibitem {21}A. A. Zyuzin and A. A. Burkov, Phys. Rev. B \textbf{83}, 195413 (2011).

\bibitem {22}M. Tahir, K. Sabeeh, and U. Schwingenschl\"{o}gl J. Appl. Phys.
\textbf{113}, 043720 (2013)

\bibitem {23}Y. L. Chen, J. G Analytis et.al, Science \textbf{325}, 178 (2009).

\bibitem {24}V. P. Gusynin, S. G. Sharapov, and J. P. Carbotte,Phys. Rev.
Lett. \textbf{98}, 157402 (2007).

\bibitem {25}V. P. Gusynin, S. G. Sharapov, and J. P. Carbotte,New J.
Phys.\textbf{11}, 095013 (2009).

\bibitem {26}Z. Li, E. A. Henniksenet al., Nat. Phys. \textbf{4}, 532 (2008).

\bibitem {27}E. J. Nicol and J. P. Carbotte,Phys. Rev. B \textbf{77}, 155409 (2008).

\bibitem {28}T. Stauber and N. M. R. Peres,J. Phys.: Condens. Matter
\textbf{20}, 055002 (2008).

\bibitem {29}M. Lasia and L. Brey, Phys. Rev. B \textbf{90}, 075417(2014)

\bibitem {30}Zhou Li and J. P. Carbotte Phys. Rev. B \textbf{88}, 045414 (2013).

\bibitem {31}Zhou Li and J. P. Carbotte,Phys.Rev.B \textbf{86}, 205425 (2012).

\bibitem {32}Zhou Li and J. P. Carbotte,Physica B \textbf{421}, 97 (2013).

\bibitem {33}L. Stille, C. J. Tabert, and E. J. Nicol, Phys. Rev. B
\textbf{86}, 195405 (2012).

\bibitem {34}V. P. Gusynin, S. G. Sharapov, and J. P. Carbotte,J. Phys.:
Condens. Matter \textbf{19}, 026222 (2007).

\bibitem {35}A, Pound, J. P. Carbotte and E. J. Nicol, Phys. Rev. B
\textbf{85}, 125422 (2012).

\bibitem {36}C. J. Tabert and E. J. Nicol, Phys. Rev. Lett. \textbf{110},
197402 (2013).

\bibitem {37}C. J. Tabert and E. J. Nicol, Phys. Rev. B \textbf{88}, 085434 (2013).

\bibitem {38}C. X. Liu, X. L. Qi, H. Zhang, X. Dai, Z. Fang, and S. C. Zhang
Phys. Rev. B \textbf{82}, 045122 (2010).

\bibitem {39}G. D. Mahan Many-Particle Physics, Third Edition, Kluwer
Academic/Plenum Publishers (2000).

\bibitem {40}N. P. Butch, K. Kirshenbaum, P. Syers, A. B. Sushkov, G. S.
Jenkins, H. D. Drew, and J. Paglione Phys. Rev B \textbf{81}, 241301(R) (2010).

\bibitem {41}J. Wang, H. Mabuchi, and X-L. Qi, Phys. Rev. B \textbf{88},
195127 (2013).

\bibitem {42}L. Fu, Phys. Rev. Lett. \textbf{103}, 266801 (2009).

\bibitem {43}E.V. Repin, V.S. Stolyarov, T. Cren, C. Brun, S.I. Bozhko, L.V.
Yashina, D. Roditchev, and I.S. Burmistrov, arXiv:1408.6960 (2014).
\end{thebibliography}
\end{document}